\documentclass[11pt]{article}
\usepackage[margin=1in]{geometry}
\usepackage{mathtools,amssymb,amsthm}
\usepackage{algorithm,algpseudocode}
\usepackage{tikz}
\usetikzlibrary{arrows.meta}
\usepackage{caption}
\usepackage{natbib}
\usepackage{microtype}
\usepackage[hidelinks]{hyperref}
\allowdisplaybreaks
\newtheorem{theorem}{Theorem}
\newtheorem{lemma}{Lemma}
\newtheorem{corollary}{Corollary}
\newtheorem{definition}{Definition}
\newtheorem{assumption}{Assumption}
\newtheorem{proposition}{Proposition}

\newcommand{\size}[1]{\ensuremath{|#1|}}
\newcommand{\ceil}[1]{\ensuremath{\lceil#1\rceil}}
\newcommand{\floor}[1]{\ensuremath{\lfloor#1\rfloor}}
\newcommand{\lrc}[1]{\ensuremath{\{#1\}}}
\newcommand{\EE}[1]{\ensuremath{\mathbb{E}[#1]}}
\def\E{\mathcal{E}}
\def\G{\mathcal{G}}
\def\I{\mathcal{I}}
\def\O{\mathcal{O}}
\def\R{\mathcal{R}}
\def\T{\mathcal{T}}
\def\OPT{\mathrm{opt}}

\title{Approximation and Inapproximability of Non-Preemptive Dial-a-Ride}
\author{Jingyang Zhao\thanks{Kyung Hee University. Email: \texttt{jingyangzhao1020@gmail.com}.}
\and
Mingyu Xiao\thanks{University of Electronic Science and Technology of China. Email: \texttt{myxiao@gmail.com}.}}
\date{}

\begin{document}
\maketitle
\begin{abstract}
The non-preemptive dial-a-ride problem (DaRP) asks for a minimum-cost route of a capacitated vehicle that serves a collection of pickup-and-delivery requests, without allowing objects to be transferred or temporarily unloaded.
We study this problem and the closely related minimum-ratio Steiner $k$-Forest (S$k$F) in general symmetric metrics on $n$ locations.
We develop a randomized polynomial-time algorithm with expected approximation ratio $\mathcal{O}(n^{1/4}\log^{11}n)$ for the minimum-ratio S$k$F.
Our algorithm combines a deterministic approximation on hierarchically well-separated trees with randomized metric embeddings.
Using this algorithm, we obtain expected $\mathcal{O}(n^{1/4}\log^{12}n)$-approximations for both the non-preemptive DaRP and the S$k$F.
These guarantees improve on the previous $n^{0.3+\varepsilon}$-approximation ratios for the S$k$F and the non-preemptive DaRP, where $\varepsilon>0$ is any fixed constant.
Complementing these algorithmic results, we establish the first superconstant inapproximability bounds for the non-preemptive DaRP.
Under the Exponential Time Hypothesis (ETH), no deterministic polynomial-time algorithm achieves an $x^{1/(\log\log x)^c}$-approximation for some constant $c>0$, where $x$ denotes either the number of locations $n$ or the vehicle capacity $k$.
Under the quantitative Gap-ETH, we rule out every deterministic polynomial-time $x^{o(1)}$-approximation.
These bounds hold even on unit-edge stars with $n=(2+o(1))k$.
\end{abstract}

\section{Introduction}\label{sec:intro}
The dial-a-ride problem (DaRP) is a classical pickup-and-delivery routing problem in transportation and logistics~\citep{journal:psaraftis:TS:80,journal:cordeau:TRB:03,journal:cordeau:OR:06,book:toth:BOOK:14}.
Its applications include door-to-door transportation for elderly passengers and people with disabilities~\citep{journal:cordeau:AOR:07}, as well as urban courier and package delivery services~\citep{journal:berbeglia:EJOR:10}.
The problem also arises in modern ride-sharing services, where passengers with similar trips share vehicles~\citep{journal:luo:VLDB:23,journal:zhang:MS:23,journal:gaul:EJOR:25}.
A service provider must assign requests to vehicles and determine pickup and delivery sequences so that all requests are served without violating vehicle capacity constraints.
These applications motivate both the single-vehicle setting and multi-vehicle extensions with different depots.

In the DaRP, the input consists of a symmetric metric $(V,c)$ on $n=\size{V}$ locations, a depot $r\in V$, a vehicle of capacity $k\in\mathbb Z_{>0}$, and a multiset of $m$ unit requests.
Each request $(s_j,t_j)$ represents an object that must be picked up at its source $s_j$ and delivered to its destination $t_j$.
The vehicle starts and ends at the depot and carries at most $k$ objects at any time.
The goal is to serve all requests while minimizing the total travel cost.

Two variants are commonly considered.
In the \emph{preemptive} variant, the vehicle may leave an object at an intermediate location and pick it up again later.
In the \emph{non-preemptive} variant, each object remains on the vehicle from pickup to delivery.
We study the non-preemptive DaRP, allowing repeated requests with multiplicities specified in binary.

\subsection{Related Work}
\paragraph{DaRP.}
When $k=1$, the non-preemptive DaRP is the stacker-crane problem (SCP), for which \citet{journal:frederickson:SICOMP:78} obtained a $\frac95$-approximation.
The DaRP also contains the metric traveling salesman problem (TSP) as the special case in which every source is the depot and $k=m$, and is therefore NP-hard~\citep{book:traub:BOOK:24}.
For the metric TSP, \citet{report:christofides:TECHREP:76} and \citet{journal:serdyukov:UPRAV:78} obtained the classical $\frac32$-approximation, and \citet{conference:karlin:IPCO:23,journal:karlin:OR:24} improved the ratio to $\frac32-10^{-36}$ with a deterministic algorithm.

For the non-preemptive DaRP, \citet{conference:charikar:FOCS:98} obtained an $\O(\sqrt k\log n\log\log n)$-approximation.
Their algorithm, combined with the tree embedding of \citet{journal:fakcharoenphol:JCSS:04}, achieves an $\O(\sqrt k\log n)$-approximation.

The Steiner $k$-Forest (S$k$F) seeks a minimum-cost forest connecting at least $k$ specified request pairs in a connected undirected graph with nonnegative edge costs.
Its minimum-ratio variant seeks a tree connecting between one and $k$ selected requests at minimum cost per selected request.
In this variant, only selected requests count toward the limit, even if the tree connects other requests, and the tree need not contain the depot.
\citet{journal:gupta:TALG:10} showed that any $\alpha$-approximation for the minimum-ratio S$k$F yields an $\O(\alpha\log^2 n)$-approximation for the non-preemptive DaRP.
They obtained an $\O(\min\lrc{\sqrt k,\sqrt n})$-approximation for the minimum-ratio S$k$F, and hence an $\O(\min\lrc{\sqrt k,\sqrt n}\log^2n)$-approximation for the non-preemptive DaRP.
They also obtained an $\O(\min\lrc{\sqrt n\log k,\sqrt k})$-approximation for the S$k$F, improving on the $\O(\min\lrc{n^{2/3},\sqrt m}\log n)$-approximation obtained by \citet{journal:segev:ALGCA:10} via a Lagrangian relaxation.
Note that any $\alpha$-approximation for the S$k$F also yields an $\O(\alpha)$-approximation for the minimum-ratio S$k$F~\citep{journal:gupta:TALG:10}

For every fixed $\varepsilon>0$, \citet{journal:dinitz:TALG:17} later obtained $\O(n^{(7-4\sqrt2)/3+\varepsilon})$-approximations for the S$k$F with unit edge costs and the non-preemptive DaRP on metrics induced by graphs with unit edge costs.
For positive integer edge costs at most $n^\beta$, where $\beta\ge0$ is fixed, their approximation ratio becomes $\O(n^{(1+\beta)(7-4\sqrt2)/3+\varepsilon})$.
\citet{misc:zhao:arXiv:26} obtained an $\O(\min\lrc{\sqrt{k}\log m,\sqrt{m/k},\sqrt{n\log n}})$-approximation for the non-preemptive DaRP in general metrics.
Recently, for every fixed $\varepsilon>0$, \citet{misc:chlamtac:arXiv:26} obtained a randomized $n^{0.3+\varepsilon}$-approximation for the S$k$F, and thus the same approximation ratio for the non-preemptive DaRP in general metrics.

For the preemptive DaRP, the algorithm of \citet{conference:charikar:FOCS:98}, combined with the embedding of \citet{journal:fakcharoenphol:JCSS:04}, achieves an $\O(\log n)$-approximation.
However, the optimal non-preemptive cost can exceed the optimal preemptive cost by a factor of $\Omega(n^{1/3})$ in general metrics~\citep{conference:charikar:FOCS:98} and $\Omega(n^{1/8}/\log^3n)$ even in the Euclidean plane~\citep{journal:gupta:TALG:10}.

For the non-preemptive DaRP with multiple vehicles starting at possibly different depots, \citet{journal:luo:VLDB:23} developed a hierarchical grouping algorithm with approximation ratio $\O(\sqrt{k}\log m)$ for minimizing the total travel cost.
For the preemptive variant, \citet{journal:gortz:TALG:15} studied the makespan objective, which minimizes the maximum completion time of the vehicles, including their return to their depots.
They obtained an $\O(\log^3 n)$-approximation whose schedule preempts each object at most once.
When $k=\infty$, they obtained an $\O(\log t)$-approximation, where $t\le n$ is the number of distinct depots.
For metrics induced by graphs excluding a fixed minor, their approximation ratios improve to $\O(\log^2 n)$ and $\O(1)$, respectively.

There are fewer results on inapproximability.
For the preemptive DaRP, \citet{conference:gortz:APPROX:06} proved a lower bound of $\Omega(\min\lrc{\log^{1/4-\varepsilon}n,k^{1-\varepsilon}})$ for every fixed $\varepsilon>0$, unless all problems in NP can be solved by randomized algorithms with expected running time $\O(n^{\text{poly}\log n})$.
For the non-preemptive variant, the previously known lower bound was only constant, inherited from the metric TSP~\citep{journal:karpinski:JCSS:15}.

\paragraph{S$k$F and related problems.}
Although superconstant inapproximability for the non-preemptive DaRP had remained open, several conditional lower bounds were known for the S$k$F.
For some constant $c>0$, \citet{conference:manurangsi:STOC:17} excluded a polynomial-time $n^{1/(\log\log n)^c}$-approximation for the S$k$F under Exponential Time Hypothesis (ETH)~\citep{journal:impagliazzo:JCSS:01a}.
Under Gap-ETH, he also excluded a deterministic polynomial-time $n^{f(n)}$-approximation for every nonnegative function $f$ with $f(n)\to0$.
For the S$k$F with target $k$, \citet{conference:manurangsi:ITCS:21} ruled out an $o(\sqrt k)$-approximation in $f(k)\operatorname{poly}(n)$ time for any function $f$, under the Strongish Planted Clique Hypothesis (SPCH).
\citet{misc:chlamtac:arXiv:26} used the sum-of-squares lower bounds of \citet{conference:jones:STOC:23} to provide evidence against a polynomial-time $\O(k^{1/2-\varepsilon})$-approximation for the S$k$F, where $\varepsilon>0$ is any sufficiently small fixed constant.

These results use reductions from dense-subgraph problems to the S$k$F.
The Smallest $k$-Edge Subgraph (S$k$ES) seeks a smallest vertex set inducing at least $k$ edges.
It is a special case of the S$k$F on unit-edge stars: each graph vertex becomes a leaf, and each graph edge becomes a terminal pair.
The target $k$ is unchanged, and the forest cost equals the number of selected graph vertices~\citep{journal:dinitz:TALG:17}.
The algorithm of \citet{journal:dinitz:TALG:17} uses an extension of the S$k$ES with vertex costs and polynomially bounded edge profits.

The Densest $k$-Subgraph (D$k$S) instead seeks $k$ vertices inducing as many edges as possible.
For every fixed $\varepsilon>0$, \citet{conference:bhaskara:STOC:10} obtained an $\O(n^{1/4+\varepsilon})$-approximation for this problem, and \citet{conference:chlamtac:FOCS:12} obtained an $\O(n^{3-2\sqrt2+\varepsilon})$-approximation for the S$k$ES.
Moreover, \citet{conference:chlamtac:FOCS:12} showed that an $\alpha$-approximation for the S$k$ES yields a $\widetilde{\O}(\alpha^2)$-approximation for the D$k$S, where $\widetilde{\O}$ suppresses factors polynomial in $\log n$.
For the D$k$S, the best known approximation ratio depending only on $k$ remains $\O(k)$~\citep{conference:manurangsi:ITCS:21}.

The ETH and Gap-ETH lower bounds of \citet{conference:manurangsi:STOC:17} also apply to both the S$k$ES and the D$k$S.
The mentioned $k$-dependent inapproximability bound and sum-of-squares evidence for the S$k$F also apply to the S$k$ES.
For the D$k$S, \citet{conference:manurangsi:ITCS:21} ruled out an $o(k)$-approximation in $f(k)\operatorname{poly}(n)$ time for any function $f$, under the SPCH.

\subsection{Our Results}
We obtain improved approximation algorithms for the minimum-ratio S$k$F, the non-preemptive DaRP, and the S$k$F in general metrics.
We also establish the first superconstant inapproximability results for the non-preemptive DaRP.
Our main results are summarized below.

\begin{enumerate}
\item For the minimum-ratio S$k$F, we obtain a randomized polynomial-time algorithm with expected approximation ratio $\O(n^{1/4}\log^{11}n)$.
Our approach first develops a deterministic approximation algorithm on hierarchically well-separated trees (HSTs) and then extends it to general metrics through randomized tree embeddings.
This algorithm forms the basis of our approximations for the DaRP and the S$k$F.

\item For the non-preemptive DaRP, combining our minimum-ratio algorithm with the reductions of \citet{journal:gupta:TALG:10} yields an expected $\O(n^{1/4}\log^{13}n)$-approximation.
We further improve the reduction from the minimum-ratio S$k$F to the non-preemptive DaRP, saving a factor of $\log n$.
This leads to an expected $\O(n^{1/4}\log^{12}n)$-approximation.
The minimum-ratio algorithm also yields an expected $\O(n^{1/4}\log^{12}n)$-approximation for the S$k$F, using the Lagrangian framework of \citet{journal:segev:ALGCA:10}.

\item Complementing these algorithmic results, we prove the first superconstant inapproximability results for the non-preemptive DaRP, even on unit-edge stars with $n=(2+o(1))k$.
Our reduction builds on the birthday graph construction of \citet{conference:manurangsi:STOC:17} and uses locally computable assignment distributions to select requests deterministically.
Separately for $x=n$ and $x=k$, ETH excludes a deterministic polynomial-time $x^{1/(\log\log x)^c}$-approximation for some constant $c>0$.
Quantitative Gap-ETH~\citep{report:dinur:TECHREP:16} excludes a deterministic polynomial-time $x^{f(x)}$-approximation for every nonnegative function $f$ with $f(x)\to0$.
These lower bounds have the same asymptotic forms as the D$k$S inapproximability results of \citet{conference:manurangsi:STOC:17}.
\end{enumerate}

\section{Preliminaries}\label{sec:prelim}

\subsection{The Non-Preemptive DaRP}
Let $(V,c)$ be a symmetric metric on $n=\size{V}$ locations, where $c(u,v)$ is the travel cost from $u$ to $v$.
Let $\R=\lrc{(s_j,t_j)}_{j=1}^m$ be a multiset of $m$ unit requests.
Each request $j$ corresponds to an object that must be picked up at its source $s_j$ and delivered to its destination $t_j$.
The non-preemptive DaRP also specifies a depot $r\in V$ and a vehicle of capacity $k\in\mathbb Z_{>0}$.
A feasible route $\sigma$ starts and ends empty at $r$, serves every request, and carries at most $k$ objects at any time.
Each object remains on the vehicle from pickup to delivery.
The goal is to minimize $c(\sigma)$, the sum of the travel costs of all moves in $\sigma$.

A \emph{request type} is an ordered pair of endpoints $(u,v)\in V\times V$.
The input specifies a positive integer count for each available type, called its \emph{multiplicity}.
Each request has a distinct label, even if several requests have the same endpoints.
As in \citet{journal:gupta:TALG:10}, we allow both $m$ and $k$ to be superpolynomial in $n$.

A route specifies pickup and delivery actions and the moves between them.
Several actions may occur at one visit, and their order determines the vehicle load.
For repeated requests, an action may specify how many objects it picks up or delivers, and a closed route segment may be repeated a specified number of times.
The cost $c(\sigma)$ includes all these repetitions.
Thus, we can represent a route without listing every repeated action.
Requests whose endpoints coincide require no transport.
In routing instances, we regard them as already served and let $\R$ and $m$ refer to the remaining requests.
If no request remains, the depot-only route is optimal.

\subsection{Notation and Basic Concepts}
We view $(V,c)$ as a complete graph, assigning each edge $uv$ the cost $c(uv)=c(u,v)$.
For graph inputs, we assume that the graph is connected and use its shortest-path metric.
For a graph $C$, let $V(C)$ and $E(C)$ denote its vertex and edge sets.
Its cost $c(C)=\sum_{e\in E(C)}c(e)$ counts each edge once.
For $U\subseteq V(C)$, let $C[U]$ denote the subgraph induced by $U$.

A \emph{walk} is a sequence of vertices in which consecutive vertices are joined by an edge, and its cost counts every traversal of an edge.
A \emph{path} is a walk with no repeated vertices.

For a request multiset $S\subseteq\R$, let $U_s(S)=\lrc{s_j:j\in S}$ and $U_t(S)=\lrc{t_j:j\in S}$ denote its sets of distinct sources and destinations.
The count $\size{S}$ includes all selected copies.
A request set is \emph{simple} if it contains at most one request of each type.
A \emph{representative} of a type is one request with that type's endpoints.
We specify a selection by the number of requests selected from each type, without listing its individual labels.

For a tree $T_r$ rooted at $r$ and $U\subseteq V(T_r)$, let $\operatorname{span}_{T_r}(U)$ denote the minimal subtree containing $r$ and $U$.
Its cost counts shared edges only once.
For a request multiset $S$, we also use $\operatorname{span}_{T_r}(S)=\operatorname{span}_{T_r}(U_s(S)\cup U_t(S))$.

Logarithms have base two unless specified otherwise.

\subsection{S$k$F and Minimum-Ratio Selection}
We use the following two connection problems on the same metric and request multiset.

\begin{definition}[S$k$F]\label{def:skf}
An instance of the S$k$F consists of a symmetric metric $(V,c)$, a request multiset $\R$, and an integer target $1\le k\le m$.
The objective is to find a forest $F$ of minimum cost $c(F)$ connecting at least $k$ requests of $\R$.
A request is connected if its endpoints belong to the same component of $F$.
Every copy of a connected request type counts toward the target.
\end{definition}

\begin{definition}[Minimum-Ratio S$k$F]\label{def:min-ratio}
An instance of the \emph{minimum-ratio S$k$F} consists of a symmetric metric $(V,c)$, a nonempty request multiset $\R$ specified by type multiplicities, and a selection limit $k\in\mathbb Z_{>0}$.
A feasible \emph{connected selection} is a pair $(C,S)$, where $S\subseteq\R$ is nonempty with $\size{S}\le k$ and $C$ is a tree connecting all endpoints of $S$.
The objective is to minimize $c(C)/\size{S}$.
Only requests in $S$ count toward the limit, even if $C$ connects other requests, and $C$ need not contain the depot.
\end{definition}

The output consists of $C$ and the selected count of each type.
For $\alpha\ge1$, an $\alpha$-approximation returns a feasible connected selection whose ratio is at most $\alpha$ times the minimum ratio.
For randomized algorithms, this guarantee holds in expectation.

A request with $c(s_j,t_j)=0$ is an optimal zero-ratio selection for the minimum-ratio S$k$F and can be returned directly.
For the remaining instances, we contract zero-distance classes for metric computations.
When needed, we retain the original vertices as distinct terminal leaves attached by zero-cost edges.
For connection problems, we may orient an unordered pair according to a fixed ordering of $V$, since this preserves its connection cost.

\section{Minimum-Ratio S$k$F and Rooted Selection}\label{sec:selector}
We reduce the minimum-ratio S$k$F to rooted selection on an HST.

\subsection{Reduction to Rooted Selection}\label{sec:rooted-ratio}
We first introduce HST and the definition of rooted selection on an HST.

We first fix a vertex $r$ and require all selected endpoints to be connected to it.
Here, $r$ denotes the selection root, which need not be the routing depot.
Their rooted span is the union of their paths to $r$, with shared edges counted only once.
We try every root, including the highest vertex of an optimal unrooted span, so this requirement need not add any cost.

\begin{figure}[!t]
\centering
\caption{An HST with vertex heights and edge costs. The highlighted path has cost $4+2+2=8=h(r)-h(v)$.}
\label{fig:hst}
\smallskip
\begin{tikzpicture}[x=.78cm,y=.85cm,
  vertex/.style={circle,draw=black!65,fill=white,inner sep=1.8pt},
  terminal/.style={vertex,fill=black},
  edge/.style={draw=black!55,line width=.65pt},
  path/.style={draw=black,line width=1.15pt},
  cost/.style={fill=white,inner sep=1.5pt,font=\footnotesize},
  every node/.style={font=\footnotesize}]
\foreach \y in {0,1,2,3}
  \draw[black!15,densely dotted] (-5.8,\y)--(6.1,\y);
\node[font=\small\bfseries] at (7.2,3.6) {Height};
\foreach \y/\height in {3/8,2/4,1/2,0/0}
  \node at (7.2,\y) {$\height$};
\node[vertex] (a) at (0,3) {};
\node[above=4pt] at (a) {$r$};
\node[vertex] (b1) at (-3,2) {};
\node[vertex] (b2) at (3,2) {};
\foreach \name/\x in {c1/-4.5,c2/-1.5,c3/1.5,c4/4.5}
  \node[vertex] (\name) at (\x,1) {};
\foreach \name/\x in {v/-5.25,d2/-3.75,d3/-2.25,d4/-.75,d5/.5,d9/1.5,d6/2.5,d7/3.75,d8/5.25}
  \node[terminal] (\name) at (\x,0) {};
\node[below=5pt] at (v) {$v$};
\draw[path] (a)--node[cost]{4}(b1);
\draw[edge] (a)--node[cost]{4}(b2);
\draw[edge] (a)--node[cost]{6}(c2);
\draw[path] (b1)--node[cost]{2}(c1);
\foreach \u/\v in {b2/c3,b2/c4}
  \draw[edge] (\u)--node[cost]{2}(\v);
\draw[path] (c1)--node[cost]{2}(v);
\foreach \u/\v in {c1/d2,c2/d3,c2/d4,c3/d5,c3/d9,c3/d6,c4/d7,c4/d8}
  \draw[edge] (\u)--node[cost]{2}(\v);
\end{tikzpicture}
\end{figure}
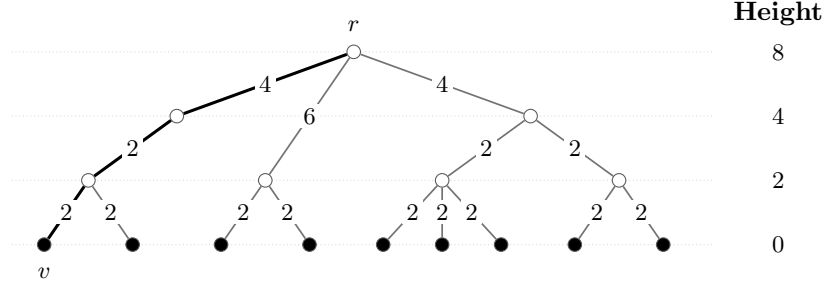

Let $T$ be a rooted HST with heights $h(v)\ge0$, where the original terminals are leaves of height zero.
For a parent $u$ and child $v$, we use $h(u)\ge2h(v)$ and $c(uv)=h(u)-h(v)$.
Let $T_r$ denote the subtree consisting of $r$ and its descendants, and let $c_T(u,v)$ denote the cost of the unique $u$--$v$ path in $T$.
Thus, $c_T(r,v)=h(r)-h(v)\le h(r)$ for every $v\in V(T_r)$. An illustration of these heights and edge costs is shown in Figure~\ref{fig:hst}.

For $U\subseteq V(T_r)$, let $g(U)=c(\operatorname{span}_{T_r}(U))$.
For a request multiset $S$, let $f(S)=g(U_s(S)\cup U_t(S))$.
These are the costs of connecting endpoints and complete requests, respectively, to $r$.
They are monotone, and $g(X\cup Y)\le g(X)+g(Y)$ because the two spans may share edges.

We also fix a target $Q$ and budget $B$.
Rather than insisting on $Q$ requests, we seek a nonempty selection of at most $Q$ requests with cost per request comparable to $B/Q$.
A smaller selection may already have the required ratio.

\begin{definition}[Rooted selection]\label{def:rooted-ratio}
A \emph{rooted selection instance} is a tuple $\I=(T_r,\R,Q,B)$, where $Q\ge1$ is an integer and $B>0$ satisfies $B\ge h(r)$.
The requests have endpoints in $T_r$, with at most $n$ distinct sources and at most $n$ distinct destinations. Endpoints may be internal vertices or coincide.
A feasible selection is a nonempty multiset $S\subseteq\R$ with $\size{S}\le Q$.
An $\alpha$-approximation returns such a selection with $f(S)/\size{S}\le\alpha B/Q$ whenever some $\R^*\subseteq\R$ has $\size{\R^*}=Q$ and $f(\R^*)\le B$.
We call $\R^*$ a \emph{comparison set}, which the algorithm does not know.
\end{definition}

\begin{lemma}\label{lem:rooted-reduction}
An $\alpha$-approximation for rooted selection yields a $4\alpha$-approximation algorithm for the minimum-ratio S$k$F on an HST whose request endpoints are original terminal leaves, and a randomized expected $\O(\alpha\log n)$-approximation algorithm in general metrics.
\end{lemma}
\begin{proof}
On an HST, we first return a zero-cost request if one exists.
Otherwise, we enumerate roots $r$, powers of two $Q\le\min\lrc{k,m}$, and budgets $B$ obtained by doubling the least positive edge cost until $B\ge c(T)$.
For each choice, we keep requests with both endpoints in $T_r$ and apply the rooted algorithm if at least $Q$ requests remain and $B\ge h(r)$.
We return the selection of minimum unrooted ratio.

Let $C^*$ be an optimal span connecting $b$ requests at cost $B^*>0$.
At its highest vertex $r$, all selected endpoints lie in $T_r$ and $B^*\ge h(r)$.
Some enumerated choice satisfies $b/2<Q\le b$ and $B^*\le B<2B^*$, so any $Q$ of these requests form a comparison set.
The returned ratio is therefore at most $\alpha B/Q\le4\alpha B^*/b$.

The embedding of \citet{journal:fakcharoenphol:JCSS:04} produces dominating HSTs with expected distortion $\O(\log n)$, so the tree guarantee extends to general metrics with an additional expected $\O(\log n)$ factor.
\end{proof}

\section{The Algorithm for Rooted Selection}\label{sec:selector-algorithm}
We present our algorithm for a rooted selection instance
$\I=(T_r,\R,Q,B)$.
Let $N=\max\lrc{\size{U_s(\R)},\size{U_t(\R)}}$
denote the size of $\I$, where $N\le n$.
As shown in Algorithm~\ref{alg:unified}, we first reduce $\I$ to simple instances $\I_\mu$ to handle repeated requests.
For each $\I_\mu$, we construct source and destination blocks, partition the requests into classes, and obtain selection candidates.
We then reduce each class to a smaller instance and recursively solve it.
Finally, we return the candidate for $\I$ with minimum ratio.

We next describe the algorithm in detail and establish an
approximation ratio of $\tilde{\O}(N^{1/4})$.

\begin{algorithm}[!t]
\caption{\textsc{MinRatio}$(T_r,\R,Q,B)$}\label{alg:unified}
\small
\begin{algorithmic}[1]
\Require A rooted selection instance $\I=(T_r, \R, Q, B)$ with $\size{\R}\ge Q\geq 1$.
\State Initialize $\T\gets\emptyset$.\label{algs1l1}

\For{each enumerated simple instance $\I_\mu=(T_r,\R_\mu,Q_\mu,B)$ with $\size{\R_\mu}\geq Q_\mu\geq 1$}\label{algs1l3}
  \State \textbf{Blocks, groups, and classes:}\label{algs1l4}
  \State Greedily construct source blocks $\E_s=\lrc{E_{s,1},E_{s,2},\ldots}$ and destination blocks $\E_t=\lrc{E_{t,1},E_{t,2},\ldots}$.\label{algs1l5}
  \State Obtain source groups $\G_s=\lrc{E_{s,1},E_{s,2},E_{s,3}\cup E_{s,4},E_{s,5}\cup\cdots\cup E_{s,8},\ldots}$, and similarly destination groups.\label{algs1l6}
  \State Form (request) classes $A=\lrc{(u,v)\in\R_\mu:u\in G_s,\ v\in G_t}$ for pairs of groups $(G_s,G_t)\in \G_s\times\G_t$.\label{algs1l7}

  \State Compute the new target $Q'$.\label{algs1l8}

  \State For each source or destination block, compute its block candidate.\Comment{Block candidates} \label{algs1l10}

  \For{each nonempty class $A$}\label{algs1l11}
  
    \State For each endpoint of $A$, compute its neighborhood candidate.\Comment{Neighborhood candidates}\label{algs1l13}

    \State \textbf{Instance reduction:}\label{algs1l14}
    \State Move some endpoints of $A$ toward the root, then remove certain endpoints and their incident requests.\label{algs1l15}
    \State Let $A'$ be the remaining requests, preserving their original identities.\label{algs1l16}   
    \State If $\size{A'}\ge Q'\geq 1$, recurse on $(T_r,A',Q',B)$ and retain the returned candidate.\Comment{Recursive candidates}\label{algs1l18}
  \EndFor\label{algs1l19}
  \State For each candidate computed for $\I_\mu$, restore its  endpoints if necessary, take $\mu$ copies of each selected request, and add the result to $\T$.
\EndFor\label{algs1l20}
\State \Return the candidate in $\T$ with minimum ratio.\label{algs1l21}
\end{algorithmic}
\end{algorithm}

\subsection{Reduction to Simple Instances}\label{sec:binary-wrapper}
Fix $\ell=c_0\log n$ for a sufficiently large constant $c_0$.
A rooted selection instance may contain repeated requests.
We show that, at an $\O(\ell)$ loss in the approximation factor, it suffices to work on simple instances.

\begin{lemma}\label{lem:binary-scale}
For a rooted selection instance $\I=(T_r,\R,Q,B)$, we can compute in polynomial time at most $\ell$ simple instances $\I_\mu=(T_r,\R_\mu,Q_\mu,B)$, where $Q_\mu=\ceil{Q/(4\ell\mu)}$ and each type in $\R_\mu$ has at least $\mu$ copies in $\R$.
If $\I$ admits a comparison set, at least one simple instance also admits a comparison set.
Moreover, any $\alpha$-approximate selection for this instance $\I_\mu$ can be converted into a $4\ell\alpha$-approximate selection for $\I$.
\end{lemma}
\begin{proof}
Enumerate the powers of two $\mu$ in $[\max\lrc{1,Q/(4n^2)},Q]$, yielding at most $\ell$ choices.
For each $\mu$, we construct $\I_\mu$ by keeping one representative of each request type with at least $\mu$ copies in $\R$, and set its target to $Q_\mu$.
Suppose that $\I$ admits a comparison set $\R^*$ of $Q$ requests.
We first discard from $\R^*$ all types contributing fewer than $Q/(2n^2)$ copies.
Since there are at most $n^2$ types, at most $Q/2$ requests are discarded.

For each remaining type, we round its number of copies down to the nearest power of two. All these values are among the enumerated choices of $\mu$.
Since at least $Q/2$ requests remain, some value $\mu$ accounts for at least $Q/(2\ell)$ requests.
Each type with this rounded value has fewer than $2\mu$ copies in $\R^*$, so there are at least $\ceil{Q/(4\ell\mu)}=Q_\mu$ such types.
These types all belong to $\R_\mu$, and their rooted span costs at most $B$.
Thus, $\I_\mu$ also admits a comparison set.
Finally, given an $\alpha$-approximate selection for $\I_\mu$, we take $\mu$ copies of each selected request type. This preserves the rooted span and selects at most $\mu Q_\mu\le Q$ requests. Since $\mu Q_\mu\ge Q/(4\ell)$, the resulting selection for $\I$ has ratio at most $\alpha B/(\mu Q_\mu)\le4\ell\alpha B/Q$.
\end{proof}

\subsection{Constructing Blocks, Groups, and Request Classes}\label{sec:local}\label{sec:normalization}
Fix an enumerated value $\mu$ and its instance $\I_\mu=(T_r,\R_\mu,Q_\mu,B)$.
We introduce the constructions in lines~\ref{algs1l4}--\ref{algs1l6}. 

\paragraph{Constructing source and destination blocks.}
We first construct two lists of source and destination blocks, denoted by
$\E_s=\lrc{E_{s,1},E_{s,2},\ldots}$ and $\E_t=\lrc{E_{t,1},E_{t,2},\ldots}$, respectively. Each block has rooted cost at most $\varrho B$, where $\varrho=\tilde{O}(N^{1/4})$ is chosen later.
In particular, for sources and destinations separately, we repeatedly
select as many remaining endpoints as possible within budget $\varrho B$, record them as the next block, and remove them from consideration.
Since $B\ge h(r)$, every endpoint can be connected individually within this budget, so all endpoints are eventually assigned to blocks.
Note that the blocks in each list are pairwise disjoint, but source and destination blocks may share tree vertices.

\paragraph{Grouping blocks and classifying requests.}
We next group the blocks in $\E_s$ and $\E_t$ separately using the index ranges $\lrc{1}$, $\lrc{2}$, $\lrc{3,4}$, $\lrc{5,6,7,8}$, and so on, truncating the last range.
A \emph{group} is the union of the blocks in one range.
Let $\G_s$ and $\G_t$ denote the resulting sets of source and destination groups.
Within each group, the largest block index is at most twice the smallest.

For each pair $(G_s,G_t)\in\G_s\times\G_t$, we define a request class $A=\lrc{(u,v)\in\R_\mu:u\in G_s,\ v\in G_t}$, as shown in lines~\ref{algs1l6}--\ref{algs1l7}.
These classes partition $\R_\mu$, and there are at most $\ell^2$ classes.
If $\I_\mu$ admits a comparison set $\R_\mu^*$, then some class $A^*$ satisfies $\size{\R_\mu^*\cap A^*}\ge Q_\mu^*\coloneqq\ceil{Q_\mu/\ell^2}$.

\begin{figure}[t]
\centering
\caption{Illustration of blocks, groups, and request classes.}
\begingroup
\linespread{1}\selectfont
\begin{tikzpicture}[x=1cm,y=1cm,
  every node/.style={font=\fontsize{9}{10.5}\selectfont},
  block/.style={draw=black!70,fill=white,rounded corners=1pt,
    minimum width=.49cm,minimum height=.42cm,inner sep=1pt},
  group/.style={draw=black!65,densely dashed,rounded corners=2pt},
  selected/.style={group,draw=black,line width=.9pt,fill=black!8},
  request/.style={draw=black,line width=.7pt}]
\path[use as bounding box] (-.1,0) rectangle (12.05,4.25);
\node[anchor=east,align=right] at (1.9,3.1) {source blocks};
\node[anchor=east,align=right] at (1.9,1.2) {destination blocks};
\foreach \yy/\labely/\last in {3.1/3.69/13,1.2/.61/16} {
  \foreach \lo/\hi/\lab in {1/1/{1},2/2/{2},3/4/{3,4},5/8/{5,\ldots,8},9/\last/{9,\ldots,\last}} {
    \pgfmathsetmacro{\xl}{2.3+.62*(\lo-1)-.30}
    \pgfmathsetmacro{\xr}{2.3+.62*(\hi-1)+.30}
    \pgfmathsetmacro{\xm}{(\xl+\xr)/2}
    \draw[group] (\xl,\yy-.32) rectangle (\xr,\yy+.32);
    \node at (\xm,\labely) {$\{\lab\}$};
  }
}
\draw[selected] (4.48,2.78) rectangle (6.94,3.42);
\draw[selected] (6.96,.88) rectangle (11.90,1.52);
\node at (5.71,4.06) {$G_s$};
\node at (9.43,.24) {$G_t$};
\foreach \j in {1,...,13} {
  \pgfmathsetmacro{\xx}{2.3+.62*(\j-1)}
  \node[block] (s\j) at (\xx,3.1) {$\j$};
}
\foreach \j in {1,...,16} {
  \pgfmathsetmacro{\xx}{2.3+.62*(\j-1)}
  \node[block] (t\j) at (\xx,1.2) {$\j$};
}
\draw[request] (s5.south)--(t10.north);
\draw[request] (s7.south)--(t14.north);
\draw[request] (s8.south)--(t16.north);
\node[fill=white,inner sep=3pt] at (8.05,2.15) {class $A$};
\end{tikzpicture}
\endgroup
\label{fig:selector-blocks}
\end{figure}
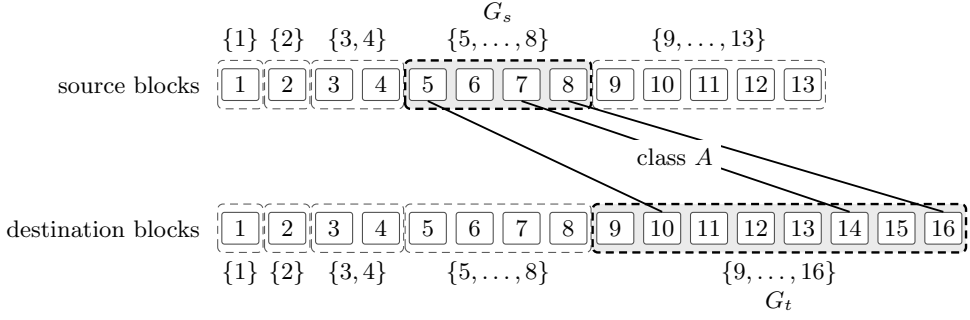

An illustration of the blocks, groups, and request classes is shown in Figure~\ref{fig:selector-blocks}.

\begin{lemma}\label{lem:block-construction}
All greedy source and destination blocks can be constructed in polynomial time.
\end{lemma}
\begin{proof}
Each greedy step is equivalent to finding a rooted subtree of cost at most $\varrho B$ containing as many remaining endpoints as possible.
This can be done in polynomial time via a tree dynamic program (DP).
Since every remaining endpoint can be connected to the root within budget $\varrho B$, each step removes at least one endpoint.
\end{proof}

\subsection{Constructing Block and Neighborhood Candidates}\label{sec:direct-candidates}
We use the following operation to construct both kinds of candidates.
Given a rooted subtree $P$, we consider requests with at least one endpoint in $P$ and select a nonempty set $S$ of at most $Q_\mu$ such requests minimizing $c(P\cup\operatorname{span}_{T_r}(S))/\size{S}$. 
We call this operation \emph{completion from $P$}.

\paragraph{Block candidates.}
For each source or destination block, let $P$ be its rooted span.
We apply completion from $P$ to the requests in $\R_\mu$ and call the resulting selection its \emph{block candidate}.

A block candidate provides a good approximation if its block is incident to sufficiently many requests of a comparison set, since these requests can be completed at an additional cost of at most $B$.

\paragraph{Neighborhood candidates.}
For each source or destination endpoint $v$ of a class $A$, we consider its distinct neighbors in $A$, namely, the opposite
endpoints of the requests in $A$ incident to $v$.

For each possible number $t$ of neighbors, we select $t$ neighbors with minimum rooted cost and let $P_t$ be their rooted span.
We then apply completion from $P_t$ using only requests of $A$.
Among the resulting selections, we keep the one with minimum ratio as the \emph{neighborhood candidate} of $v$.

These candidates help preserve comparison requests during instance reduction.
If too many comparison requests are lost, we will show that a neighborhood candidate already provides a good approximation.
Otherwise, sufficiently many comparison requests remain for recursion.

Next, we show that the above candidates can be constructed in polynomial time.
For each $t$, a minimum-rooted-cost set of $t$ neighbors can be found by a tree DP, as in the construction of source and destination blocks.
It then remains to show that completion can be done in polynomial time.

\begin{lemma}\label{lem:complete}
Completion from a rooted subtree $P$ can be performed in polynomial time.
Moreover, its returned nonempty selection $S$ satisfies $\size{S}\le Q_\mu$ and $\frac{f(S)}{\size{S}}\le\frac{c(P\cup\operatorname{span}_{T_r}(J))}{|J|}\le\frac{c(P)+f(J)}{\size{J}}$ for every nonempty eligible request set $J$ with $\size{J}\le Q_\mu$.
\end{lemma}
\begin{proof}
For each eligible request, choose an endpoint in $P$ and place one item at its other endpoint.
Set the edge costs in $P$ to zero, obtaining $c'$, and add $c(P)$ separately.

Let $W_v(b)$ be the minimum modified cost of connecting $b$ selected items in the subtree rooted at $v$ to $v$, excluding its parent edge.
If $m_v$ items lie at $v$, initialize $W_v(b)=0$ for $0\le b\le m_v$ and $+\infty$ otherwise.
We process the tree bottom-up, incorporating each child $u$ by
\begin{equation}\label{eq:count}
 W_v^{\mathrm{new}}(b)=
 \min_{0\le j\le b}
 \lrc{W_v(b-j)+W_u(j)+c'(vu)\mathbf 1_{\lrc{j>0}}}.
\end{equation}
We consider counts only up to the smaller of $Q_\mu$ and the number of eligible requests, so the DP runs in polynomial time.

For every nonempty eligible $J$ with $\size{J}\le Q_\mu$, the DP can select $J$, so $c(P)+W_r(\size{J})\le c(P\cup\operatorname{span}_{T_r}(J))$.
We return a selection minimizing $(c(P)+W_r(b))/b$ over all feasible positive counts.
If the returned selection contains $b$ requests, its rooted cost is at most $c(P)+W_r(b)$.
The claimed bound follows since $c(P\cup\operatorname{span}_{T_r}(J))\le c(P)+f(J)$.
\end{proof}

A direct property is that requests are sparse between block pairs unless a good block candidate is found.

\begin{lemma}
\label{lem:sparse-blocks}
Either a block candidate for $\I_\mu$ has ratio at most $8\varrho\ell^2B/Q_\mu$, or every source--destination block pair contains fewer than $Q_\mu/(4\ell^2)$ requests.
\end{lemma}
\begin{proof}
Suppose some source--destination block pair contains at least $Q_\mu/(4\ell^2)$ requests, and let $J$ consist of $\ceil{Q_\mu/(4\ell^2)}$ of them.
Their endpoints lie in two blocks whose rooted spans have total cost at most $2\varrho B$.
By Lemma~\ref{lem:complete}, completion from the source block yields a candidate for $\I_\mu$ with ratio at most $\frac{2\varrho B}{|J|}\le \frac{8\varrho\ell^2B}{Q_\mu}$.
\end{proof}

\subsection{Reducing the Instance}\label{sec:recursive-selection}
We process each nonempty request class $A$ separately.
Let $N'$ be an integer with $0\le N'<N$, to be chosen in the analysis.
We construct a recursive instance with at most $N'$ distinct sources and at most $N'$ distinct destinations.
Its target is $Q'=\ceil{Q_\mu^*/2}$, corresponding to half the comparison requests guaranteed in $A^*$.

\paragraph{Moving endpoints.}
We first move nearby endpoints to common ancestors, so that several endpoints may occupy the same tree vertex.
Let $\delta>0$ be a parameter to be chosen.
For every endpoint $v$ with $h(v)\le\delta$, we move $v$ to its highest ancestor in $T_r$ of height at most $\delta$.
All other endpoints remain unchanged.
Let $\pi$ denote this map and $\pi A$ the requests after their endpoints move.
Each endpoint moves toward $r$ by at most $\delta$.
Thus, restoring both endpoints of a returned request increases its rooted cost by at most $2\delta$.
We allow an additional cost of $\varrho B/Q_\mu^*$ per returned request and therefore set $\delta=\varrho B/(2Q_\mu^*)$.

\paragraph{Removing endpoints.} 
We view $\pi A$ as a bipartite graph whose two vertex sets consist of its distinct source and destination endpoints, respectively.
Each request corresponds to a separate edge, even if several requests have the same endpoints.
We compute all degrees and retain up to $N'$ source vertices and up to $N'$ destination vertices of largest degree.
We then remove the remaining vertices and their incident requests.

\paragraph{Recursive candidates.}
Let $A'$ be the remaining request set.
If $\size{A'}\ge Q'$, we recursively solve
$(T_r,A',Q',B)$ as in line~\ref{algs1l18}.
If the call returns a selection, restore the original endpoints of the returned requests, take $\mu$ copies of each, and add the resulting selection for $\I$ to $\T$.

\subsection{Analysis of the Instance Reduction}\label{sec:reduction-analysis}
Suppose $\I_\mu$ admits a comparison set $\R_\mu^*$. Recall that $A^*$ contains at least $Q_\mu^*=\ceil{Q_\mu/\ell^2}$ requests of $\R_\mu^*$.
Fix any set $I\subseteq\R_\mu^*\cap A^*$ of $Q_\mu^*$ requests.
We show that either a block or neighborhood candidate for $\I_\mu$ has ratio at most $8\varrho\ell^2B/Q_\mu$, or the recursive instance corresponding to $A^*$ retains at least $Q'=\ceil{Q_\mu^*/2}$ requests of $I$.

For the sake of analysis, we make the following assumption.
\begin{assumption}\label{ass:no-direct-candidate}
Every block and neighborhood candidate for $\I_\mu$
has ratio greater than $8\varrho\ell^2B/Q_\mu$.
\end{assumption}

We first establish bounds on $A^*$ and the number of distinct endpoints of $I$, and then analyze which requests of $I$ survive the endpoint reduction.

Let $b_s,b_t$ be the largest block indices in the source and destination groups defining $A^*$.

\begin{lemma}\label{lem:class-bounds}
Under Assumption~\ref{ass:no-direct-candidate}, the class $A^*$ satisfies $\size{A^*}<\frac{Q_\mu b_sb_t}{4\ell^2}\le \frac{Q_\mu^*b_sb_t}{4}$, $g(U_i(A^*))\le b_i\varrho B$ for $i\in\lrc{s,t}$, and $b_s,b_t>4$.
\end{lemma}
\begin{proof}
By Lemma~\ref{lem:sparse-blocks}, each block pair contains fewer than $Q_\mu/(4\ell^2)$ requests. Since $A^*$ involves at most $b_sb_t$ such pairs, the first bound holds.
For each $i\in\lrc{s,t}$, all side-$i$ endpoints of $A^*$ belong to the corresponding group, which consists of at most $b_i$ blocks.
Since each block has rooted cost at most $\varrho B$, the union of their rooted spans has cost at most $b_i\varrho B$ by construction.
Thus, the second bound holds.

To prove $b_s,b_t>4$, consider the requests in $I$.
For each $i\in\lrc{s,t}$, their side-$i$ endpoints belong to at most $b_i$ blocks. Thus, some block is incident to at least $\frac{Q_\mu^*}{b_i}$ requests of $I$.
These requests have rooted cost at most $B$, and the block has rooted cost at most $\varrho B$.
By Lemma~\ref{lem:complete}, its block candidate for $\I_\mu$ has ratio at most $\frac{(\varrho+1)B}{Q_\mu^*/b_i}\le\frac{2\varrho b_iB}{Q_\mu^*}$.
If $b_i\le4$, this is at most $\frac{8\varrho B}{Q_\mu^*}\le\frac{8\varrho\ell^2B}{Q_\mu}$, contrary to our assumption.
Thus, $b_s,b_t>4$.
\end{proof}

Since $b_s,b_t>4$, the group construction ensures that at least $b_i/2$ blocks precede the side-$i$ group defining $A^*$, even if the last group is truncated.
We next use the greedy choice of these earlier blocks to bound the number of distinct endpoints of requests from $I$.
Let $n_i=\size{U_i(\R_\mu)}\le N$ for $i\in\lrc{s,t}$.

\begin{lemma}\label{lem:norm}
Under Assumption~\ref{ass:no-direct-candidate}, the set $I$ satisfies $f(I)\le B$ and $\size{U_i(I)}\le\frac{4n_i}{\varrho b_i}$ for $i\in\lrc{s,t}$.
\end{lemma}
\begin{proof}
Since $I\subseteq\R_\mu^*$, we have $f(I)\le B$.
For $i\in\lrc{s,t}$, consider any block $E_{\mathrm{pre}}$ preceding the side-$i$ group defining $A^*$.
All endpoints in $U_i(I)$ were available when $E_{\mathrm{pre}}$ was selected, and $g(U_i(I))\le B$.
The greedy block construction therefore implies $\size{E_{\mathrm{pre}}}\ge\size{U_i(I)}$.
We show that $\size{E_{\mathrm{pre}}}>\frac{\varrho\size{U_i(I)}}{2}$.

First, for each $1\le t\le\size{E_{\mathrm{pre}}}$, some $t$-element subset $E'_t\subseteq E_{\mathrm{pre}}$ satisfies $g(E'_t)\le\frac{(t-1)g(E_{\mathrm{pre}})}{\size{E_{\mathrm{pre}}}}+B$.

To see this, order the endpoints of $E_{\mathrm{pre}}$ by a depth-first traversal. Starting from each endpoint, take the next $t$ consecutive endpoints in cyclic order, obtaining $\size{E_{\mathrm{pre}}}$ subsets of size $t$.

For each edge $e$ in the rooted span of $E_{\mathrm{pre}}$, the endpoints below $e$ form a consecutive interval in the depth-first order.
The rooted span of a subset contains $e$ if and only if the subset includes an endpoint below $e$.
Thus, the number of these subsets whose rooted spans contain $e$ is at most $t-1$ plus the number of endpoints below $e$.

Averaging their rooted costs, we obtain $\frac{(t-1)g(E_{\mathrm{pre}})+\sum_{v\in E_{\mathrm{pre}}}c_T(r,v)}{\size{E_{\mathrm{pre}}}}\le\frac{(t-1)g(E_{\mathrm{pre}})}{\size{E_{\mathrm{pre}}}}+B$, where the inequality follows from $c_T(r,v)\le h(r)\le B$.
Thus, at least one subset has rooted cost at most $\frac{(t-1)g(E_{\mathrm{pre}})}{\size{E_{\mathrm{pre}}}}+B$.

Now suppose $\size{E_{\mathrm{pre}}}\le\frac{\varrho\size{U_i(I)}}{2}$ and set $t=\size{E_{\mathrm{pre}}}-\size{U_i(I)}+1$.
The subset $E'_t$ then satisfies $g(E'_t\cup U_i(I))\le\frac{(\size{E_{\mathrm{pre}}}-\size{U_i(I)})\varrho B}{\size{E_{\mathrm{pre}}}}+2B\le\varrho B$.
Since $E_{\mathrm{pre}}$ and $U_i(I)$ are disjoint, $E'_t\cup U_i(I)$ contains $\size{E_{\mathrm{pre}}}+1$ available endpoints, contrary to the greedy choice of $E_{\mathrm{pre}}$.
Thus, every preceding block contains more than $\frac{\varrho\size{U_i(I)}}{2}$ endpoints.

Since $b_i>4$ by Lemma~\ref{lem:class-bounds}, at least $b_i/2$ such blocks precede the side-$i$ group defining $A^*$, and they are pairwise disjoint. Thus, $n_i>\frac{b_i}{2}\cdot\frac{\varrho\size{U_i(I)}}{2}=\frac{\varrho b_i\size{U_i(I)}}{4}$.
\end{proof}

Lemma~\ref{lem:class-bounds} bounds the requests and rooted costs in $A^*$, while Lemma~\ref{lem:norm} bounds the numbers of distinct sources and destinations of the fixed set $I$.
We now use these bounds to analyze which requests of $I$ remain after their endpoints move and some vertices are removed.

\paragraph{Bounds after moving and removing endpoints.}
We first bound the number of distinct endpoints after moving them and the cost of restoring them.

\begin{lemma}\label{lem:projection}
For every $v\in V(T_r)$, $U\subseteq V(T_r)$, and request set $S$ with endpoints in $T_r$, we have $c_T(v,\pi(v))\le\delta$, $f(\pi S)\le f(S)$, and $\size{\pi(U)}\le1+\frac{2g(U)}{\delta}$.
Moreover, restoring both endpoints costs at most $2\delta=\varrho B/Q_\mu^*$ per selected request.
\end{lemma}
\begin{proof}
Each endpoint moves along its path to $r$ by a distance of at most $\delta$.
Thus, the rooted span after moving the endpoints is contained in the original rooted span.
Restoring a selected request requires adding two paths, each of cost at most $\delta$.

To prove the last bound, for each $u\in\pi(U)\setminus\lrc{r}$, consider its incoming edge $pu$ in $T_r$.
If $h(u)\le\delta$, the definition of $\pi$ implies $h(p)>\delta$, so $c(pu)=h(p)-h(u)\ge h(p)/2>\delta/2$.
Otherwise, $h(u)>\delta$ and $c(pu)\ge h(u)>\delta$.
These edges are distinct and belong to the rooted span of $U$.
Their total cost is at most $g(U)$, so $\frac{\delta}{2}(\size{\pi(U)}-1)\le g(U)$ and then $\size{\pi(U)}\le1+\frac{2g(U)}{\delta}$.
\end{proof}

We next consider a vertex removed from $\pi A$.

\begin{lemma}\label{lem:removed-degree}
Every vertex removed from $\pi A$ has degree at most $\size{A}/(N'+1)$ before removal.
\end{lemma}
\begin{proof}
If a source vertex $u$ is removed, the $N'$ retained source vertices each have degree at least that of $u$.
Since the sum of all source degrees is $\size{A}$, we have $(N'+1)\deg_{\pi A}(u)\le\size{A}$. The same argument applies to destination vertices.
\end{proof}

\paragraph{Bounds from neighborhood candidates.}
The endpoint removal may discard requests of $I$.
We show that if too many such requests are discarded, some neighborhood candidate for $\I_\mu$ has a small ratio.

\begin{lemma}\label{lem:progress}
Under Assumption~\ref{ass:no-direct-candidate}, if $\varrho\ge1$ satisfies $\varrho^4(N'+1)^2\ge 2^{18}N^3$, then the recursive instance corresponding to $A^*$ retains at least $Q'$ requests of $I$ with total rooted cost at most $B$, and has at most $N'$ distinct sources and at most $N'$ distinct destinations.
\end{lemma}
\begin{proof}
Let $\gamma$ be the minimum ratio among all block and neighborhood candidates for $\I_\mu$.
By Assumption~\ref{ass:no-direct-candidate}, $\gamma>8\varrho\ell^2B/Q_\mu$.
Since $Q'=\ceil{Q_\mu^*/2}$, it suffices to show that at most $Q_\mu^*/4$ requests of $I$ are lost through source removal, and the same bound holds for destination removal.

By Lemma~\ref{lem:norm}, $I$ has at most $r_i=\frac{4n_i}{\varrho b_i}$ distinct endpoints on side $i\in\lrc{s,t}$.
Suppose more than $Q_\mu^*/4$ requests of $I$ are lost because the images of their side-$i$ endpoints are removed, and let $j$ denote the other side.
We first remove those requests whose side-$j$ endpoint is incident to fewer than $\frac{Q_\mu^*}{8r_j}$ requests of $I$.
Since $I$ has at most $r_j$ distinct endpoints on side $j$, fewer than $Q_\mu^*/8$ requests are removed.
Thus, more than $Q_\mu^*/8$ lost requests remain.
They are incident to at most $r_i$ distinct endpoints on side $i$.
Consequently, some original endpoint $v$ whose image $\pi(v)$ is removed is incident to more than $\frac{Q_\mu^*}{8r_i}$ of these requests.
Let $Y$ consist of their opposite endpoints.
Since $A^*$ is simple before its endpoints move, these endpoints are distinct, so $\size{Y}>\frac{Q_\mu^*}{8r_i}$.
Each vertex in $Y$ is incident to at least $\frac{Q_\mu^*}{8r_j}$ requests of $I$.
Moreover, $Y\subseteq U_j(I)$, and Hence, $g(Y)\le f(I)\le B$.

Recall that the neighborhood candidate of $v$ considers a minimum-rooted-cost set $X_t$ of $t$ distinct neighbors for every possible $t$. For the analysis only, consider one of these sets maximizing $2Bt-\size{Y}g(X_t)$. Since $X_t$ has minimum rooted cost among all sets of $t$ neighbors, the chosen set also maximizes $2B\size{X}-\size{Y}g(X)$ over all nonempty sets $X$ of neighbors of $v$.
In particular, comparing $X_t$ with $X_t\cup Y$, we have $2B\size{Y\setminus X_t} \le\size{Y}(g(X_t\cup Y)-g(X_t))\le\size{Y}B$, where the last inequality follows from $g(Y)\le B$. 
Thus, $\size{X_t\cap Y}\ge\size{Y}/2$.
Similarly, comparing $X_t$ with $Y$, we have $2B\size{X_t}-\size{Y}g(X_t)\ge2B\size{Y}-\size{Y}g(Y)\ge B\size{Y}$. Thus, $g(X_t)+B\le2B\size{X_t}/\size{Y} \le2\deg_{A^*}(v)B/\size{Y}$, since $A^*$ is simple and $X_t$ contains at most $\deg_{A^*}(v)$ neighbors.

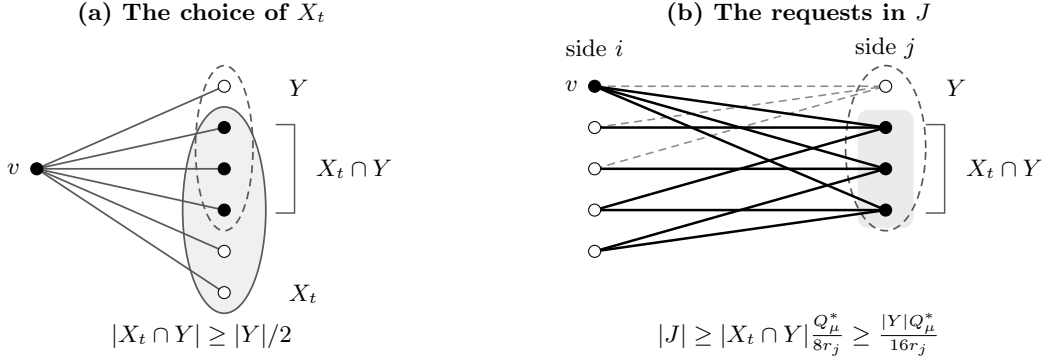
\begin{figure}[t]
\centering
\caption{Illustration for the neighborhood analysis.
(a) The chosen set $X_t$ contains at least half of $Y$.
(b) The requests of $I$ whose side-$j$ endpoints lie in $X_t\cap Y$ form the set $J$.}
\begingroup
\linespread{1}\selectfont
\begin{tikzpicture}[x=1cm,y=.78cm,
  every node/.style={font=\fontsize{9}{10.5}\selectfont},
  endpoint/.style={circle,draw=black,fill=white,inner sep=1.6pt},
  marked endpoint/.style={endpoint,fill=black},
  request/.style={draw=black!65,line width=.65pt},
  chosen request/.style={draw=black,line width=.95pt},
  other request/.style={draw=black!45,densely dashed,line width=.6pt},
  Y set/.style={draw=black!65,densely dashed,line width=.65pt},
  X set/.style={draw=black!65,fill=black!6,line width=.65pt}]
\path[use as bounding box] (-.15,-.55) rectangle (14.55,5.80);

\node[font=\footnotesize\bfseries] at (2.80,5.53) {(a) The choice of $X_t$};
\draw[X set] (3.10,2.20) ellipse [x radius=.55cm,y radius=1.365cm];
\draw[Y set] (3.10,3.25) ellipse [x radius=.38cm,y radius=1.092cm];
\coordinate (v) at (.62,2.90);
\foreach \j/\yy in {1/4.30,2/3.60,3/2.90,4/2.20,5/1.50,6/.80} {
  \coordinate (n\j) at (3.10,\yy);
  \draw[request] (v)--(n\j);
}
\node[marked endpoint,label=left:{$v$}] at (v) {};
\foreach \j in {1,5,6} {
  \node[endpoint] at (n\j) {};
}
\foreach \j in {2,3,4} {
  \node[marked endpoint] at (n\j) {};
}
\node[anchor=west] at (3.82,4.28) {$Y$};
\node[anchor=west] at (3.82,.78) {$X_t$};
\draw[black!65,line width=.55pt] (3.78,3.65)--(4.03,3.65)
  --(4.03,2.15)--(3.78,2.15);
\node[anchor=west] at (4.18,2.90) {$X_t\cap Y$};
\node[anchor=base] at (2.80,-.03) {$\size{X_t\cap Y}\ge\size{Y}/2$};

\node[font=\footnotesize\bfseries] at (10.70,5.53) {(b) The requests in $J$};
\node at (8.00,4.95) {side $i$};
\node at (11.85,4.95) {side $j$};
\fill[black!8,rounded corners=4pt] (11.48,1.90) rectangle (12.22,3.90);
\draw[Y set] (11.85,3.25) ellipse [x radius=.53cm,y radius=1.092cm];
\foreach \j/\yy in {1/4.30,2/3.60,3/2.90,4/2.20,5/1.50} {
  \coordinate (a\j) at (8.00,\yy);
}
\foreach \j/\yy in {1/4.30,2/3.60,3/2.90,4/2.20} {
  \coordinate (b\j) at (11.85,\yy);
}
\foreach \j in {1,2,3} {
  \draw[other request] (a\j)--(b1);
}
\foreach \u/\w in {1/2,2/2,4/2,1/3,3/3,5/3,1/4,4/4,5/4} {
  \draw[chosen request] (a\u)--(b\w);
}
\node[marked endpoint,label=left:{$v$}] at (a1) {};
\foreach \j in {2,3,4,5} {
  \node[endpoint] at (a\j) {};
}
\node[endpoint] at (b1) {};
\foreach \j in {2,3,4} {
  \node[marked endpoint] at (b\j) {};
}
\node[anchor=west] at (12.52,4.28) {$Y$};
\draw[black!65,line width=.55pt] (12.38,3.65)--(12.63,3.65)
  --(12.63,2.15)--(12.38,2.15);
\node[anchor=west] at (12.78,2.90) {$X_t\cap Y$};
\node[anchor=base] at (10.70,-.03)
  {$\size{J}\ge\size{X_t\cap Y}\frac{Q_\mu^*}{8r_j}
    \ge\frac{\size{Y}Q_\mu^*}{16r_j}$};
\end{tikzpicture}
\endgroup
\label{fig:selector-neighborhood}
\end{figure}

Let $J$ consist of the requests of $I$ whose side-$j$ endpoint belongs to $X_t\cap Y$.
Each vertex of $X_t\cap Y$ is incident to at least $\frac{Q_\mu^*}{8r_j}$ requests of $I$.
Since these vertices are all on side $j$, their incident request sets are disjoint.
Then, since $\size{X_t\cap Y}\ge\size{Y}/2$, we have $\size{J}\ge\frac{\size{Y}Q_\mu^*}{16r_j}$.
Moreover, $\size{J}\le Q_\mu^*\le Q_\mu$ and $f(J)\le B$.
All requests of $J$ belong to $A^*$ and have an endpoint in the rooted span of $X_t$.
An illustration of the choice of $X_t$ and the requests in $J$ is shown in Figure~\ref{fig:selector-neighborhood}.
By Lemma~\ref{lem:complete}, the neighborhood candidate of $v$ for $\I_\mu$ has ratio at most $\frac{g(X_t)+B}{\size{J}}$.
Thus, 
\begin{equation}\label{eq:neighbor-completion}
 \gamma\le\frac{g(X_t)+B}{\size{J}}\leq\frac{32\deg_{A^*}(v)r_jB}{\size{Y}^2Q_\mu^*}<\frac{2048\size{A^*}r_i^2r_jB}{(N'+1)(Q_\mu^*)^3}<\frac{2^{15}N^3B}{\varrho^3(N'+1)b_i(Q_\mu^*)^2},
\end{equation}
where the second inequality follows from $\size{J}\ge\frac{\size{Y}Q_\mu^*}{16r_j}$, the third from $\size{Y}>\frac{Q_\mu^*}{8r_i}$ and $\deg_{A^*}(v)\le\frac{\size{A^*}}{N'+1}$ by Lemma~\ref{lem:removed-degree}, and the last from $n_i,n_j\le N$, $r_i=\frac{4n_i}{\varrho b_i}$, and $\size{A^*}<\frac{Q_\mu^*b_sb_t}{4}$ by Lemma~\ref{lem:class-bounds}.

By Lemmas~\ref{lem:class-bounds} and~\ref{lem:projection}, together with $\delta=\frac{\varrho B}{2Q_\mu^*}$, we obtain $\size{U_i(\pi A^*)}\le1+4b_iQ_\mu^*\le5b_iQ_\mu^*$.
Since $\pi(v)$ is removed, $\pi A^*$ must have at least $N'+1$ distinct endpoints on side $i$, and thus, $b_i\ge\frac{N'+1}{5Q_\mu^*}$.
Then, by \eqref{eq:neighbor-completion},
\[
\gamma\leq \frac{5\cdot2^{15}N^3B}{\varrho^3(N'+1)^2Q_\mu^*}\leq \frac{5\varrho B}{8Q_\mu^*}<\frac{\varrho B}{Q_\mu^*},
\]
where the second inequality uses $\varrho^4(N'+1)^2\ge2^{18}N^3$ by assumption.
This contradicts $\gamma>8\varrho\ell^2B/Q_\mu\ge\frac{8\varrho B}{Q_\mu^*}$.

Thus, at most $Q_\mu^*/4$ requests of $I$ are lost through source removal, and at most $Q_\mu^*/4$ through destination removal.
Consequently, at least $Q'=\ceil{Q_\mu^*/2}$ requests of $I$ remain, with rooted cost at most $B$ by Lemma~\ref{lem:projection}.
\end{proof}

\subsection{Approximation Analysis}\label{sec:oracle-analysis}

\begin{theorem}\label{thm:oracle}
Algorithm~\ref{alg:unified} achieves an approximation ratio of $2^{20}\ell^{10}N^{1/4}$ for rooted selection instances.
Consequently, the minimum-ratio S$k$F admits a deterministic $\O(n^{1/4}\log^{10}n)$-approximation on HSTs, and a randomized algorithm with expected approximation ratio $\O(n^{1/4}\log^{11}n)$ in general metrics.
\end{theorem}
\begin{proof}
Let $\rho(N)=2^{20}\ell^{10}N^{1/4}$. We set
\begin{equation}\label{eq:parameter-choice}
N'=\floor{\frac{N}{2^{16}\ell^{12}}},\qquad
\varrho=\min\lrc{2^j:j\in\mathbb Z_{\ge0},\ 2^{4j}(N'+1)^2\ge2^{18}N^3}.
\end{equation}
Thus, the condition of Lemma~\ref{lem:progress} holds.
Since $N'+1>\frac{N}{2^{16}\ell^{12}}$, the choice of $\varrho$ implies $\varrho<2(\frac{2^{18}N^3}{(N'+1)^2})^{1/4}<2^{14}\ell^6N^{1/4}$.
We recompute $N$, $N'$, and $\varrho$ in each recursive call, while $n$ and $\ell$ remain fixed.

We prove the ratio bound by induction on $N$.
Assume that $\I$ admits a comparison set and that the bound holds for every smaller instance.
By Lemma~\ref{lem:binary-scale}, some simple instance $\I_\mu$ admits a comparison set.
Recall that $\frac{Q}{\mu Q_\mu}\le4\ell$, $\frac{Q}{\mu Q_\mu^*}\le4\ell^3$, and $\frac{Q}{\mu Q'}\le8\ell^3$.
If a block or neighborhood candidate for $\I_\mu$ has ratio at most $8\varrho\ell^2B/Q_\mu$, we use that candidate.
Otherwise, Lemma~\ref{lem:progress} ensures that the recursive instance corresponding to $A^*$ admits a comparison set and has size at most $N'<N$.

In the first case, we take $\mu$ copies of each selected request to obtain a selection for $\I$ with ratio at most $\frac{8\varrho\ell^2B}{\mu Q_\mu}\le\frac{32\varrho\ell^3B}{Q}\le\frac{\rho(N)B}{Q}$.

In the second case, the induction hypothesis bounds the ratio of the selection returned for $(T_r,A',Q',B)$ by $\frac{\rho(N')B}{Q'}$.
We then restore its endpoints to obtain a selection for $\I_\mu$, adding at most $\frac{\varrho B}{Q_\mu^*}$ per request by Lemma~\ref{lem:projection}.
After we take $\mu$ copies, the resulting selection for $\I$ has ratio at most $(8\ell^3\rho(N')+4\ell^3\varrho)\frac{B}{Q}$.
Since $N'\le\frac{N}{2^{16}\ell^{12}}$, we have $8\ell^3\rho(N')\le\rho(N)/2$.
Moreover, the bound on $\varrho$ implies $4\ell^3\varrho<\rho(N)/2$.
Thus, the claimed ratio follows.
Note that if $N'=0$, no nonempty recursive instance exists, so Lemma~\ref{lem:progress} ensures the first case.

For feasibility, each completion selects at most $Q_\mu$ representatives, and each recursive selection contains at most $Q'\le Q_\mu$ requests.
If several requests coincide after their endpoints move, we assign each returned copy to a distinct recorded request with those endpoints before restoring its original endpoints.
We then take $\mu$ copies of each selected representative, so at most $\mu Q_\mu\le Q$ requests of $\I$ are selected.

For the running time, each call considers at most $\ell$ simple instances and $\ell^2$ classes per instance.
Each class produces at most one recursive instance of size at most $N'$.
Thus, the sum of the sizes of all recursive instances in one call is at most $\ell^3N'\le\frac{N}{2^{16}\ell^9}<N/2$.
The total number of nonempty recursive calls is therefore $\O(N)$.
Each local computation takes polynomial time, and request multiplicities are stored as binary counts.

Since $N\le n$ and $\ell=\O(\log n)$, the rooted approximation ratio is $\O(n^{1/4}\log^{10}n)$.
By Lemma~\ref{lem:rooted-reduction}, this implies the claimed expected approximation ratio in general metrics.
\end{proof}

\section{Algorithms for the Non-Preemptive DaRP and Steiner $k$-Forest}\label{sec:application}
We now derive approximation algorithms for the non-preemptive DaRP and the S$k$F from our algorithm for the minimum-ratio S$k$F.

\subsection{The Non-Preemptive DaRP}\label{sec:darp}

\citet{journal:gupta:TALG:10} showed the following useful reductions.

\begin{lemma}[{\citet[Theorems~3.3 and 3.9]{journal:gupta:TALG:10}}]\label{lem:ghnr-reductions}
An $\alpha$-approximation algorithm for the minimum-ratio S$k$F yields an $\O(\alpha\log^2 m)$-approximation algorithm for the non-preemptive DaRP with $m$ requests. Moreover, an $\alpha$-approximation algorithm for the non-preemptive DaRP with $\O(n^4)$ requests yields an $\O(\alpha)$-approximation algorithm for the non-preemptive DaRP with arbitrarily many requests.
\end{lemma}

Combining Lemma~\ref{lem:ghnr-reductions} with Lemma~\ref{lem:rooted-reduction} and Theorem~\ref{thm:oracle}, we obtain the following corollary.

\begin{corollary}\label{cor:darp-ghnr}
The non-preemptive DaRP admits a deterministic $\O(n^{1/4}\log^{12}n)$-approximation on HSTs, and a randomized algorithm with expected approximation ratio $\O(n^{1/4}\log^{13}n)$ in general metrics.
\end{corollary}

The randomization in general metrics arises from the HST embedding, which incurs an additional expected $\O(\log n)$ factor.
We provide a better reduction from the minimum-ratio S$k$F to the non-preemptive DaRP.

\begin{theorem}\label{thm:main-upper}
An $\alpha$-approximation algorithm for the minimum-ratio S$k$F yields an $\O(\alpha\log m)$-approximation algorithm for the non-preemptive DaRP with $m$ requests.
In particular, the non-preemptive DaRP admits a deterministic $\O(n^{1/4}\log^{11}n)$-approximation on HSTs, and a randomized algorithm with expected approximation ratio $\O(n^{1/4}\log^{12}n)$ in general metrics.
\end{theorem}
\begin{proof}
For the non-preemptive DaRP, let $\OPT=c(\sigma)$ be the cost of an optimal route $\sigma$.
For any nonempty multiset $\R'$ of $u$ remaining requests, let $z$ be the minimum ratio of a connected selection of at most $k$ requests from $\R'$.
We first show that $zu\le2\OPT$.

We may assume that $z>0$.
We follow $\sigma$ but omit service actions for requests outside $\R'$.
For $0<w<kz$, we divide the route into segments of cost $w$ and a possible final segment of cost less than $w$.
For every sufficiently small $\varepsilon>0$, we can
choose $w\in(kz-\varepsilon,kz)$ such that no interior
cut coincides with a service action.
Since only finitely many values in this interval
are forbidden, such a choice always exists.

Each segment must contain fewer than $k$ requests whose pickups and deliveries both occur within it.
Otherwise, the segment connects $k$ requests at cost less than $kz$, contrary to the definition of $z$.
Similarly, if a segment contains $b$ such requests, its cost is at least $zb$.
Thus, at most $\OPT/z$ requests are completed within individual segments.
Every other request is on the vehicle at some interior cut.
Recall that $k$ denotes the vehicle capacity, so at most $k$ objects are on board at each cut.
Since there are at most $\OPT/w$ cuts, we have $u\le\frac{\OPT}{z}+\frac{k\OPT}{w}$.
Since $w\in(kz-\varepsilon,kz)$ holds for arbitrarily small $\varepsilon>0$, we obtain $zu\le2\OPT$.

We now repeatedly apply the $\alpha$-approximation algorithm for the minimum-ratio S$k$F to the remaining requests and remove each selected set.
When $u$ requests remain, suppose it selects $1\le s\le k$ requests connected by a tree $T_s$.
Then $\frac{c(T_s)}{s}\le\alpha z\le\frac{2\alpha\OPT}{u}$.
Since $\frac{s}{u}\le\sum_{j=u-s+1}^u\frac{1}{j}$, summing over all iterations shows that the total cost of the selected trees is at most $2\alpha H_m\OPT$, where $H_m=\sum_{j=1}^m1/j$.

For each selected tree, we choose one selected source as its starting vertex.
We first traverse its doubled edges in a closed walk to pick up all selected objects, then repeat the walk to deliver them.
This costs at most four times the tree cost.
We connect these starting vertices and the depot by a doubled minimum spanning tree, and insert each tree's service walk at its starting vertex.
The connecting walk costs at most $2\OPT$, since an optimal route visits all these vertices.
The resulting feasible non-preemptive route has cost at most $(8\alpha H_m+2)\OPT=\O(\alpha\log m)\OPT$.
Thus, the first claim of the theorem holds.
The second claim then follows by further using Lemma~\ref{lem:ghnr-reductions}, Theorem~\ref{thm:oracle}, and randomized HST embeddings.
\end{proof}

\subsection{Steiner $k$-Forest}\label{sec:skf}
\citet{journal:gupta:TALG:10} showed that a greedy argument similar to set cover converts an $\alpha$-approximation for the minimum-ratio S$k$F into an $\O(\alpha\log k)$-approximation for the S$k$F.
They further observed that the Lagrangian framework of \citet{journal:segev:ALGCA:10} can replace $\log k$ by $\log n$.
We also consider using this framework.

\begin{theorem}\label{thm:skf}
The S$k$F admits a deterministic $\O(n^{1/4}\log^{11}n)$-approximation algorithm on HSTs, and a randomized polynomial-time algorithm with expected approximation ratio $\O(n^{1/4}\log^{12}n)$ in general metrics.
\end{theorem}
\begin{proof}
We use the Lagrangian framework of \citet{journal:segev:ALGCA:10}, which also handles repeated requests.
For each set of remaining request types and budget $B$, the framework requires an algorithm that returns a tree of cost at most $\alpha B$ whose cost per connected request type is at most $\alpha$ times the minimum among trees of cost at most $B$. 
(This algorithm is required only for simple instances, as request multiplicities are handled as profits by the Lagrangian framework.)
The framework then yields an $\O(\alpha\log n)$-approximation for the S$k$F.

We verify that our algorithm satisfies these requirements on HSTs.
For each set of remaining request types and budget $B$, we run our minimum-ratio algorithm on one representative per type with every possible request limit, and retain the trees of cost at most $\alpha B$.
Suppose $C^*$ has minimum cost per connected type among trees of cost at most $B$ and connects $q$ types.
The call with limit $q$ returns a tree $C$ and a nonempty selection $S$ with $\size{S}\le q$ and $c(C)/\size{S}\le\alpha c(C^*)/q$.
Thus, $c(C)\le\alpha B$, and its cost per connected type is at most $c(C)/\size{S}$.
We therefore choose a retained tree of minimum cost per connected type to satisfy both requirements.

By Lemma~\ref{lem:rooted-reduction} and Theorem~\ref{thm:oracle}, we may let $\alpha=\O(n^{1/4}\log^{10}n)$ on HSTs, which proves the first bound.
Then, randomized HST embeddings incur an additional expected $\O(\log n)$ factor, proving the second bound.
\end{proof}

\section{Inapproximability under ETH and Gap-ETH}\label{sec:hardness}

We reduce Boolean the satisfiability problem (SAT) to the non-preemptive DaRP on a unit-edge star.
We first transform the input formula into a formula with a constant satisfiability gap and locally computable assignment distributions.
We then construct the birthday graph of \citet{conference:manurangsi:STOC:17} from this formula and use the local distributions to select edges that become requests.
In YES, a small collection of satisfying assignments covers all requests and yields a low-cost route.
In NO, the density bound of \citet{conference:manurangsi:STOC:17}, together with the route comparison in Theorem~\ref{thm:main-upper}, implies a high route cost.
We then choose the parameters to establish the lower bounds under ETH and Gap-ETH.

\subsection{The Source Formula and Local Assignment Probabilities}\label{sec:source}

For a Boolean formula, SAT asks whether some assignment of truth values makes it true.
An assignment specifies a value in $\lrc{0,1}$ for each variable, where $1$ denotes true and $0$ denotes false.
A \emph{literal} is either a variable $x$ or its negation $\neg x$.
A \emph{clause} is a disjunction of literals, such as $x\lor\neg y\lor z$.
An assignment satisfies a clause if at least one of its literals is true, and violates it otherwise.
A formula in conjunctive normal form (CNF) is a conjunction of clauses and is satisfied when every clause is satisfied.
In a 3-CNF formula, each clause contains three literals.

The original input is a \emph{sparse} 3-CNF formula with $q$ clauses and $\Theta(q)$ variables.
The sparse form of ETH excludes deterministic $2^{o(q)}$-time algorithms for deciding whether such a formula is satisfiable~\citep{journal:impagliazzo:JCSS:01b}.
We first apply the following form of Dinur's probabilistically checkable proof (PCP) theorem to obtain $\psi$.

\begin{lemma}[{\citet[Theorem~2.4]{conference:manurangsi:STOC:17}}]\label{lem:pcp-source}
There are constants $\varepsilon_0,d_0>0$ and a deterministic polynomial-time reduction from any 3-CNF formula with $q$ clauses to a 3-CNF formula $\psi$ with $\O(q\operatorname{polylog}q)$ clauses.
Each variable of $\psi$ occurs in at most $d_0$ clauses.
If the input is satisfiable, then $\psi$ is satisfiable.
Otherwise, every assignment violates at least an $\varepsilon_0$ fraction of its clauses.
\end{lemma}

The formula $\psi$ already has bounded occurrence and a constant gap.
\citet{conference:manurangsi:STOC:17} used such a formula to construct a birthday graph and prove approximation hardness for the D$k$S.
In YES, the graph contains a large clique, whereas in NO, every vertex set of the same size induces few edges.
However, this guarantee alone does not suffice for the DaRP, since a low-cost route must serve \emph{all} retained requests in YES.

We will represent request endpoints by assignments to subsets of the variables of a new formula $\phi$.
In the construction below, a satisfying assignment of $\phi$ \emph{covers} a request if it agrees with both endpoint assignments.
We seek computable request probabilities such that, in YES, each equals the probability that an assignment drawn from a common distribution of satisfying assignments of $\phi$ covers that request.
We can then retain requests with sufficiently large probabilities so that a small collection of satisfying assignments covers all retained requests in YES.
The challenge is to compute these probabilities deterministically without finding a satisfying assignment.
The next proposition constructs $\phi$ from $\psi$ with these properties, while preserving a constant gap and bounded occurrence.
We call the Boolean variables of $\phi$ \emph{positions}.

\begin{proposition}[*]\label{prop:gauge-source}
Fix any constant $\varepsilon_0>0$ and let $\psi$ be a 3-CNF formula with $q_0$ clauses.
Suppose that $\psi$ is satisfiable in YES, while every assignment violates at least an $\varepsilon_0$ fraction of its clauses in NO.
A deterministic polynomial-time algorithm constructs a 3-CNF formula $\phi$ with variable set $V_\phi$ of size $s=\Theta(q_0)$ and $\O(q_0)$ clauses.
Each variable occurs in $\O(1)$ clauses.
In YES, $\phi$ is satisfiable.
In NO, every assignment violates at least an $\varepsilon_1$ fraction of the clauses of $\phi$.
The constant $\varepsilon_1>0$ depends only on $\varepsilon_0$, and not on the instance sizes $q_0$ or $s$.

The algorithm also outputs a simple undirected graph $J$ with vertex set $V_\phi$ and maximum degree $\O(1)$.
For every $W\subseteq V_\phi$ such that every component of $J[W]$ has at most four vertices, there is a distribution $\mu_W$ on Boolean assignments to $W$.
A deterministic algorithm lists all assignments of positive $\mu_W$-probability and their exact rational probabilities in $6^{\O(\size W)}s^{\O(1)}$ time.
Every assignment $\sigma$ with $\mu_W(\sigma)>0$ satisfies all clauses of $\phi$ whose variables lie in $W$, on both YES and NO inputs.
In YES, all these distributions are restrictions of one distribution on complete satisfying assignments of $\phi$.
\end{proposition}

For the ETH reduction, we apply the proposition to $\psi$ from Lemma~\ref{lem:pcp-source}.
The reduction computes the local distributions needed for the request construction below, on both YES and NO inputs.
The common distribution is used to analyze the YES cost and need not be computed or sampled.

We take the minimum number of disjoint copies of the entire formula $\phi$ needed to obtain at least $q$ variables.
We retain the notation $\phi$, $V_\phi$, and $s=\size{V_\phi}$ for the resulting formula.
The additional copies introduce fewer than $q$ variables, so the size bounds above imply $q\le s\le Cq(\log q)^{a_0}$ for fixed constants $C>0$ and an integer $a_0\ge1$.
We take the disjoint union of the corresponding copies of $J$ and, for each $W\subseteq V_\phi$, use the product of the local distributions on its intersections with the copies.
This preserves the constant gap, bounded occurrence, and all the stated distribution guarantees.

We next construct the birthday graph and requests from the resulting formula $\phi$.

\subsection{The Birthday Graph and Request Construction}\label{sec:hard-parameters}

\paragraph{Birthday graph.}
We use the construction of \citet{conference:manurangsi:STOC:17}.
Set $\ell=\floor{s/g}$, where $g_0\le g\le s^{1/8}$ and $g_0$ is a sufficiently large fixed constant.
For sufficiently large $s$, this ensures $s^{7/8}/2\le\ell\le s/g_0$.
A vertex is a pair $(S,\sigma)$, where $S\subseteq V_\phi$ contains $\ell$ positions and $\sigma:S\to\lrc{0,1}$ assigns their Boolean values.
Vertices with distinct position sets $S,T$ are adjacent when their assignments agree on $S\cap T$ and jointly satisfy every clause whose variables lie in $S\cup T$.
There are $K=\binom{s}{\ell}$ position sets and $N_{\mathrm B}=2^\ell K$ vertices.
The standard estimates $(s/\ell)^\ell\le K\le(es/\ell)^\ell$ imply $\log N_{\mathrm B}=\Theta(\frac{s\log g}{g})$, uniformly over the chosen range of $g$.

Each satisfying assignment of $\phi$ determines one vertex for every position set through its restriction to that set.
These $K$ vertices then form a clique.
We set the vehicle capacity to $k=\binom K2$, the number of edges in such a clique.
Thus, in YES, the birthday graph contains $k$ edges on only $K$ vertices, whereas in NO, the density bound below prevents such a concentration of edges.

\begin{lemma}[{\citet[Theorem~3.1]{conference:manurangsi:STOC:17}}]\label{lem:birthday-density}
Let $\phi$ be a 3-CNF formula on $s$ variables, each occurring in at most $d$ clauses, such that every assignment violates at least an $\varepsilon_1$ fraction of clauses, where $\varepsilon_1,d>0$ are fixed.
There is a constant $\eta>0$, depending only on $\varepsilon_1,d$, such that, for every integer $\ell$ with $s^{3/4}/\eta\le\ell\le s/2$, every $K$-vertex set in its birthday graph induces at most $\binom K2 2^{-\eta\ell^4/s^3}$ edges, where $K=\binom s\ell$.
\end{lemma}

In NO, the formula $\phi$ satisfies the assumptions of Lemma~\ref{lem:birthday-density} with gap $\varepsilon_1$ and a fixed occurrence bound $d$.
We choose a sufficiently small positive rational constant $\eta$ for which Lemma~\ref{lem:birthday-density} holds, and set
\begin{equation}\label{eq:hard-density-parameter}
 \Lambda=2^{\floor{\eta\ell^4/(2s^3)}}.
\end{equation}
For sufficiently large $s$, our range of $\ell$ satisfies the lemma's requirement $s^{3/4}/\eta\le\ell\le s/2$.
In NO, every $K$-vertex set therefore induces at most $\binom K2/\Lambda^2$ edges.

\paragraph{Request construction.}
We use the graph $J$ to choose pairs of position sets on which Proposition~\ref{prop:gauge-source} computes probabilities for birthday edges.
Call distinct $\ell$-sets $S,T$ \emph{good} if every component of $J[S\cup T]$ has at most four positions.
For such a pair, the proposition computes $\mu_{S\cup T}$.
Each assignment $\sigma:S\cup T\to\{0,1\}$ with $\mu_{S\cup T}(\sigma)>0$ satisfies all clauses whose variables lie entirely in $S\cup T$ by Proposition~\ref{prop:gauge-source}, so its restrictions define the edge between $(S,\sigma|_S)$ and $(T,\sigma|_T)$.
We assign this edge probability $\mu_{S\cup T}(\sigma)$.
The two endpoint assignments uniquely determine $\sigma$ on $S\cup T$, so distinct listed assignments define distinct edges.
In YES, this is exactly the probability that a random satisfying assignment $\tau$ from the common distribution covers the edge.

We must keep enough pairs of position sets for the YES and NO costs to remain well separated.
Let $\zeta$ be the proportion of good pairs among all $\binom K2=k$ pairs of distinct $\ell$-element position sets.
Let $D$ be a fixed upper bound on the maximum degree of $J$.
Since $J$ has bounded degree, the following lemma provides a lower bound on the proportion of good pairs.

\begin{lemma}[*]\label{lem:retained-mass}
For a constant $C_D>0$ depending only on $D$ and sufficiently large $s$, we have $\zeta\ge\frac14\exp(-C_D\ell^5/s^4)$ throughout the chosen range.
\end{lemma}

\paragraph{Selecting requests.}
For each good pair $S,T$, the probabilities assigned to its edges sum to $\sum_{\sigma:S\cup T\to\{0,1\}}\mu_{S\cup T}(\sigma)=1$.
This equality holds on both YES and NO inputs, since it requires only the local distribution.
Their total over all good pairs is therefore $\zeta k$.
We next choose which edges to retain.
The NO cost lower bound grows with the number of retained edges, while the YES cost upper bound decreases with their minimum probability.
We therefore seek a large product of the number of retained edges and their minimum probability.
Sort the positive probabilities as $p_1\ge\cdots\ge p_P$ and choose $m_0$ maximizing $m_0p_{m_0}$.
Retain the first $m_0$ edges, put $a_*=p_{m_0}$, and set $B_*=2+\ceil{\log_2P}$.
Since $ip_i\le m_0a_*$ for every $i$, we have $\zeta k=\sum_{i=1}^P p_i\le m_0a_*\sum_{i=1}^P\frac1i\le m_0a_*B_*$, where $\sum_{i=1}^P\frac1i\le1+\ln P\le B_*$.
Therefore,
\begin{equation}\label{eq:hard-threshold}
 a_*m_0\ge\frac{\zeta k}{B_*},\qquad B_*=\O(s),
\end{equation}
where the second bound follows from $P\le N_{\mathrm B}^2$ and $\log N_{\mathrm B}=\O(s)$.

\begin{figure}[!t]
\centering
\caption{Illustration of the conversion of retained birthday edges into DaRP requests on a unit-edge star.
In (b), solid edges represent unit-cost metric edges, while dashed arrows represent the pickup--delivery requests corresponding to the retained edges in (a), directed from earlier to later leaves in the fixed order.}
\begingroup
\linespread{1}\selectfont
\begin{tikzpicture}[x=1cm,y=1cm,
  every node/.style={font=\fontsize{9}{10.5}\selectfont},
  vertex/.style={circle,draw=black,fill=white,inner sep=1.7pt},
  graph edge/.style={draw=black!70,line width=.7pt},
  metric edge/.style={draw=black!65,line width=.7pt},
  request/.style={draw=black,densely dashed,line width=.7pt,
    shorten <=2.8pt,shorten >=3.5pt,
    -{Stealth[length=5pt,width=4.2pt]}},
  cost/.style={fill=white,inner sep=1pt}]
\path[use as bounding box] (-.15,-.35) rectangle (12.05,3.65);
\node[font=\footnotesize\bfseries] at (2.75,3.36) {(a) Retained birthday graph};
\node[font=\footnotesize\bfseries] at (9.35,3.36) {(b) Unit-edge star};

\foreach \j/\xx in {1/.5,2/2,3/3.5,4/5} {
  \coordinate (g\j) at (\xx,1.95);
}
\draw[graph edge] (g1) to[out=55,in=125] (g2);
\draw[graph edge] (g2) to[out=55,in=125] (g3);
\draw[graph edge] (g3) to[out=55,in=125] (g4);
\draw[graph edge] (g1) .. controls (1.0,3.13) and (4.5,3.13) .. (g4);
\foreach \j in {1,...,4} {
  \node[vertex] at (g\j) {};
  \node[below=6pt] at (g\j) {$(S_\j,\sigma_\j)$};
}
\draw[-{Stealth[length=4pt,width=4pt]},black!65,line width=.75pt]
  (5.65,1.95)--(6.43,1.95);

\coordinate (depot) at (9.35,.15);
\foreach \j/\xx in {1/7.1,2/8.6,3/10.1,4/11.6} {
  \coordinate (t\j) at (\xx,1.95);
  \draw[metric edge] (depot)--(t\j);
}
\draw[request] (t1) to[out=55,in=125] (t2);
\draw[request] (t2) to[out=55,in=125] (t3);
\draw[request] (t3) to[out=55,in=125] (t4);
\draw[request] (t1) .. controls (7.6,3.13) and (11.1,3.13) .. (t4);
\foreach \j in {1,...,4} {
  \node[vertex] at (t\j) {};
  \node[below=6pt,fill=white,inner sep=1pt] at (t\j) {$(S_\j,\sigma_\j)$};
}
\node[cost] at (8.20,.98) {$1$};
\node[cost] at (8.96,.98) {$1$};
\node[cost] at (9.74,.98) {$1$};
\node[cost] at (10.50,.98) {$1$};
\node[vertex,fill=black] at (depot) {};
\node[below=4pt] at (depot) {depot};
\end{tikzpicture}
\endgroup
\label{fig:hardness-requests}
\end{figure}
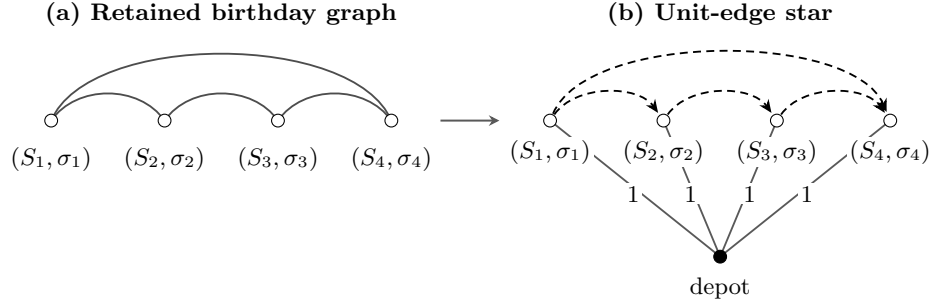

We construct a unit-edge star whose leaves correspond to the vertices incident to retained edges, with its center as the depot.
Each retained edge becomes one unit request, and we discard its probability after this choice.
Thus, this instance has $m_0$ requests.
Fix a total order of the leaves and orient every request from its earlier endpoint to its later endpoint.
As shown in the YES analysis below, each request set will be served by visiting its used leaves in this order, ensuring that every pickup precedes its delivery.
An illustration of the conversion from retained birthday edges to requests is shown in Figure~\ref{fig:hardness-requests}.

\subsection{Completeness and Soundness}\label{sec:hardness-costs}

We now compare the YES and NO route costs for the constructed requests.

\paragraph{YES.}
By Proposition~\ref{prop:gauge-source}, a random assignment from the common distribution covers each request with probability at least $a_*$.
Consider $t=\ceil{(\ln m_0+1)/a_*}$ independent assignments from this distribution.
Since $0<a_*\le1$ and $m_0\le P$, we have $a_*t\le\ln m_0+1+a_*\le\ln P+2\le B_*$.
Each request is uncovered with probability at most $(1-a_*)^t\le e^{-a_*t}$, so a union bound makes the probability that any request is uncovered at most $m_0e^{-a_*t}<1$.
Hence, some $t$ satisfying assignments cover every request.
Assign each request to one covering assignment.
Each satisfying assignment selects one birthday vertex for each of the $K$ position sets.
The requests assigned to it therefore involve at most $K$ endpoints and, since the request graph is simple, the number is at most $\binom K2=k$.
For each nonempty set, we start at the depot, visit its used leaves in the fixed order, and return to the depot.
The move from the depot to a leaf costs one, and a move between distinct leaves costs two, so the route costs twice the number of used leaves, at most $2K$.
The orientation ensures that every pickup precedes its delivery.
Since the route serves at most $k$ requests, the vehicle never carries more than $k$ objects.
We concatenate these routes to obtain $\OPT_{\rm YES}\le U:=2B_*K/a_*$.

\paragraph{NO.}
Lemma~\ref{lem:birthday-density} bounds the number of edges induced by any $K$ birthday vertices, whereas a set of at most $k$ requests may have an arbitrary number of endpoints.
We first extend this bound to such request sets.
Let $F$ be any such nonempty request set, and let $V(F)$ be its endpoint set, with $u=\size{V(F)}$.
If $u\le K$, we extend its endpoint set to $K$ vertices of the full birthday graph and apply the density bound.
If $u>K$, we choose a uniform $K$-subset of its endpoints.
Each request survives with probability $K(K-1)/(u(u-1))$, and the density bound applies to every such subset.
The expected number of surviving requests therefore satisfies
\[
 \size{F}\frac{K(K-1)}{u(u-1)}\le\frac{\binom K2}{\Lambda^2},
 \qquad\text{so}\qquad
 \size{F}\le\frac{u(u-1)}{2\Lambda^2}.
\]
In either case, $\size F\le\max\lrc{K,u}^2/(2\Lambda^2)$.
The graph is simple, so $\size F\le u^2/2$, and the selection contains at most $k=\binom K2\le K^2/2$ requests.
Hence, $\size F\le\min\lrc{K,u}^2/2$.
We multiply the two bounds and take square roots, using $\max\lrc{K,u}\min\lrc{K,u}=Ku$, to obtain $\size{F}\le\frac{K\size{V(F)}}{2\Lambda}$.
The minimum subtree spanning these $u$ leaves in the unit-edge star has cost $u$.
Hence every nonempty selection of at most $k$ requests has cost per request at least $2\Lambda/K$.
By the route comparison in the proof of Theorem~\ref{thm:main-upper}, a route serving all $m_0$ requests must therefore have cost at least $\frac{m_0}{2}\frac{2\Lambda}{K}=\frac{\Lambda m_0}{K}$.
Consequently, $\OPT_{\rm NO}\ge
L_{\rm NO}:=\Lambda m_0/K$.
Combining this with the YES upper bound
$U=2B_*K/a_*$ and \eqref{eq:hard-threshold},
the ratio between the NO lower bound and
the YES upper bound satisfies
\begin{equation}\label{eq:hardness-gap}
\Gamma:=\frac{L_{\rm NO}}{U}
=\frac{a_*m_0\Lambda}{2B_*K^2}
\ge\frac{\zeta\Lambda k}{2B_*^2 K^2}\ge\frac{\zeta\Lambda}{8B_*^2},
\end{equation}
where the inequality uses $k/K^2=(K-1)/(2K)\ge1/4$.

We next express this gap in terms of $s$ and $g$.
Since $\ell=\floor{s/g}\ge s/(2g)$ for sufficiently large $s$, the definition of $\Lambda$ implies $\log_2\Lambda\ge\frac{\eta s}{32g^4}-1$.
Lemma~\ref{lem:retained-mass} implies $\log\zeta\ge-\O(s/g^5)$, while $\log B_*=\O(\log s)$.
Thus, for some constants $c_1,c_2>0$ independent of $g$, we have
\begin{equation}\label{eq:log-gap}
 \log\Gamma\ge c_1\frac{s}{g^4}-c_2\frac{s}{g^5}-\O(\log s)
 =\Omega\left(\frac{s}{g^4}\right).
\end{equation}
We choose $g_0$ large enough that $c_2/g\le c_1/2$.
Since $g\le s^{1/8}$, we also have $s/g^4\ge s^{1/2}$, which dominates $\log s$.
Hence the last bound holds uniformly for $g_0\le g\le s^{1/8}$.

\paragraph{Locations and capacity.}\label{sec:output-parameters}
We finally arrange that $n=(2+o(1))k$, so the same cost gap proves lower bounds in either parameter.
Let $n_0\ge2$ be the number of leaves incident to retained requests.
We take $R=\ceil{2k/n_0}$ copies of the request graph, using disjoint leaf sets attached to a common depot, while keeping capacity $k$.
The new instance has $n=Rn_0+1$ locations, including the depot, and $m=Rm_0$ requests.
We can concatenate the YES routes for the copies at cost at most $RU$.
For any set $F$ of at most $k$ copied requests, let $F_i$ be its requests in copy $i$.
Each $F_i$ also contains at most $k$ requests, and the copies have disjoint endpoint sets, so
\[
 \size{F}=\sum_i\size{F_i}
 \le\frac{K}{2\Lambda}\sum_i\size{V(F_i)}
 =\frac{K\size{V(F)}}{2\Lambda}.
\]
The same ratio lower bound therefore holds for selections across copies.
By the route comparison, the $m$ copied requests have NO cost at least $RL_{\rm NO}$, even if a route mixes requests from different copies.
Thus, the copies preserve $\Gamma$ and satisfy
\begin{equation}\label{eq:output-parameters}
 2k\le n-1=Rn_0<2k+N_{\mathrm B},\qquad
 \frac{N_{\mathrm B}}{k}\le4(2\ell/s)^\ell=o(1).
\end{equation}
The first bound follows from $n_0\le N_{\mathrm B}$.
For the second, $k\ge K^2/4$ and $K\ge(s/\ell)^\ell$ imply $N_{\mathrm B}/k\le4(2\ell/s)^\ell=o(1)$, since $2\ell/s\le2/g_0<1$
and $\ell\to\infty$. Hence, $n=(2+o(1))k$.

Every leaf is incident to a request and must be visited.
Also $m_0\ge n_0/2$, so $m\ge k$.
We enumerate at most $K^2$ pairs of position sets and test which pairs are good.
For each good pair $S,T$, Proposition~\ref{prop:gauge-source} lists its exact edge probabilities in $6^{\O(\ell)}s^{\O(1)}$ time, since $\size{S\cup T}\le2\ell$.
The total is $K^2 6^{\O(\ell)}s^{\O(1)}=N_{\mathrm B}^{\O(1)}$, because $N_{\mathrm B}=2^\ell K$ and $s\le N_{\mathrm B}$.
Since $R\le2k/n_0+1$ and $m_0\le\binom{n_0}{2}$, we have $m=Rm_0\le kn_0+m_0=\O(N_{\mathrm B}^3)$.
Thus, the copied instance has $\O(N_{\mathrm B}^2)$ locations and $\O(N_{\mathrm B}^3)$ requests.
By Proposition~\ref{prop:gauge-source}, the exact probabilities can be computed and represented within $N_{\mathrm B}^{\O(1)}$ time.
Thus, sorting them, choosing the threshold, taking copies, and forming the explicit distance matrix all take $N_{\mathrm B}^{\O(1)}$ time, where the polynomial exponent is independent of $g$.

\subsection{Completing the Reductions}

Since $K\ge(s/\ell)^\ell\ge2^\ell$, we have $\log N_{\mathrm B}=\ell+\log K=\Theta(\log K)$.
Together with $k=\binom K2$ and $n=(2+o(1))k$, this implies $\log x=\Theta(\frac{s\log g}{g})$ for each output parameter $x\in\lrc{n,k}$.
We now substitute $\log N_{\mathrm B}=\Theta(\frac{s\log g}{g})$ into the polynomial bound above and use \eqref{eq:log-gap} to obtain
\begin{equation}\label{eq:hard-common-bounds}
 \text{construction time}\le N_{\mathrm B}^{\O(1)}=2^{\O(s\log g/g)},\qquad
 \Gamma=2^{\Omega(s/g^4)}.
\end{equation}
We now choose $g$ to establish the lower bounds under ETH and quantitative Gap-ETH.

\begin{theorem}\label{thm:main-lower}
For some constant $c>0$, ETH excludes a deterministic polynomial-time $x^{1/(\log\log x)^c}$-approximation for the non-preemptive DaRP, separately for $x=n$ and $x=k$.
Quantitative Gap-ETH excludes deterministic polynomial-time $x^{f(x)}$-approximation algorithms for every nonnegative function $f$ satisfying $f(x)\to0$.
These bounds hold even on unit-edge stars with $n=(2+o(1))k$.
\end{theorem}
\begin{proof}
Recall that the constructed instances have YES cost at most $RU$, NO cost at least $RL_{\rm NO}$, and ratio $\Gamma=L_{\rm NO}/U$.

\paragraph{ETH.}\label{sec:eth}
Recall that $q\le s\le Cq(\log q)^{a_0}$, so $\log s=\Theta(\log q)$.
Set $b=a_0+2$, $c=3b+1$, and $g=\ceil{(\log s)^b}$.
For sufficiently large $q$, this choice lies in the required range $g_0\le g\le s^{1/8}$.
Suppose a deterministic polynomial-time algorithm achieves the stated approximation factor, for either $x=n$ or $x=k$.
By \eqref{eq:hard-common-bounds}, the construction and this algorithm together take time at most
\[
 2^{\O\left(\frac{s\log\log s}{(\log s)^b}\right)}
 \le2^{\O\left(\frac{q\log\log q}{(\log q)^2}\right)}
 =2^{o(q)}.
\]
Moreover, the common bounds imply
\begin{equation}\label{eq:eth-size}
 \log\Gamma=\Omega\left(\frac{s}{(\log s)^{4b}}\right),\qquad
 \log x=\Theta\left(\frac{s\log\log s}{(\log s)^b}\right),\qquad
 \log\log x=\Theta(\log s).
\end{equation}
Since $c=3b+1$, we have
\[
 \log\left(x^{1/(\log\log x)^c}\right)
 =\O\left(\frac{s\log\log s}{(\log s)^{b+c}}\right)
 =\O\left(\frac{s\log\log s}{(\log s)^{4b+1}}\right)
 =o(\log\Gamma).
\]
Thus, the approximation factor is smaller than $\Gamma$ for sufficiently large $q$.
We accept exactly when the returned route has cost less than the computable threshold $\Lambda m/K=RL_{\rm NO}$.
In YES, its cost is less than $\Gamma RU=RL_{\rm NO}$.
In NO, every feasible route costs at least this threshold.
The test therefore decides satisfiability of the original input formula in $2^{o(q)}$ time, contradicting ETH.

\paragraph{Quantitative Gap-ETH.}\label{sec:gap}
We use the quantitative form of Gap-ETH~\citep{report:dinur:TECHREP:16}, with the number of clauses as the size parameter: there are fixed constants $\varepsilon_*,\tau_*>0$ such that no deterministic $\O(2^{\tau_*q})$-time algorithm distinguishes satisfiable 3-CNF formulas with $q$ clauses and $\Theta(q)$ variables from those in which every assignment violates at least an $\varepsilon_*$ fraction of clauses.
We omit Lemma~\ref{lem:pcp-source} and use the input formula itself as $\psi$, so $q_0=q$.
We then apply Proposition~\ref{prop:gauge-source} directly to $\psi$ with $\varepsilon_*$ in place of $\varepsilon_0$.
The proposition supplies bounded occurrence and the local distributions, with $s=\Theta(q)$.
Thus, the same construction and estimates apply.

Fix a supposed polynomial-time $x^{f(x)}$-approximation algorithm, where $f\ge0$ and $f(x)\to0$.
We choose $g\ge g_0$ as a sufficiently large constant integer.
For each fixed $g$, \eqref{eq:hard-common-bounds} and the output-size bound imply $\log\Gamma=\Omega(s)$ and $\log x=\Theta(s)$, with constants depending on $g$. Since $f(x)\to0$, we have $\log(x^{f(x)})=f(x)\log x=o(s)$.
The cost gap therefore exceeds the approximation factor for sufficiently large inputs.
The construction and approximation take $2^{C_A s\log g/g}$ time, where $C_A$ depends on the fixed approximation algorithm and the reduction, but not on $g$.
Since $s\le C_*q$ for a constant $C_*$ independent of $g$, and $\log g/g\to0$, we can choose $g$ sufficiently large so that the total time is at most $2^{\tau_*q/2}$.
Using the same threshold $RL_{\rm NO}$ to distinguish YES from NO then contradicts quantitative Gap-ETH.
\end{proof}

\paragraph{Randomized algorithms.}
Under the randomized versions of ETH~\citep{journal:dell:TALG:14} and quantitative Gap-ETH~\citep{conference:chalermsook:FOCS:17}, the same lower bounds hold against randomized polynomial-time approximation algorithms that return a feasible route within the claimed approximation factor with probability at least $2/3$.
The reduction is deterministic, and we accept only if the returned route is feasible and has cost below the threshold used above.
In YES, this occurs with probability at least $2/3$, whereas in NO, no feasible route can have cost below the threshold.
Thus, such an algorithm would contradict the corresponding randomized hypothesis.

\section{Omitted Proofs}
\subsection{Proof of Proposition~\ref{prop:gauge-source}}\label{app:local-source}
\begin{proof}
We first construct $\phi$ from $\psi$ and verify its size, bounded occurrence, and constant gap.
We then define $J$ and compute the local distributions on both YES and NO inputs, and show that in YES they are restrictions of one distribution of satisfying assignments of $\phi$.
An illustration of the construction and the local distributions is shown in Figure~\ref{fig:local-source}.

\begin{figure}[!t]
\centering
\caption{Illustration of the construction of $\phi$ and the local distribution $\mu_W$.}
\begingroup
\linespread{1}\selectfont
\begin{tikzpicture}[x=1cm,y=1cm,
  every node/.style={font=\fontsize{9}{10.5}\selectfont},
  stage/.style={draw=black!65,fill=white,rounded corners=2pt,
    align=center,minimum height=.95cm,inner sep=5pt},
  arrow/.style={draw=black!70,line width=.75pt,
    -{Stealth[length=4.5pt,width=4pt]}},
  note/.style={align=center,font=\fontsize{8}{9.5}\selectfont}]
\path[use as bounding box] (-.1,.05) rectangle (12.25,5.55);

\node[font=\footnotesize\bfseries] at (6.10,5.30)
  {(a) From $\psi$ to $\phi$};
\node[stage,minimum width=1.80cm] (psi) at (1.00,4.55)
  {Input formula\\$\psi$};
\node[stage,minimum width=2.60cm] (colors) at (4.10,4.55)
  {Color constraints\\on $B$};
\node[stage,minimum width=2.70cm] (permutations) at (7.85,4.55)
  {Permutation variables\\$q_j(v),p_j(e),t_j(e)$};
\node[stage,minimum width=2.10cm] (phi) at (11.05,4.55)
  {Boolean formula\\$\phi$};
\draw[arrow] (psi.east)--(colors.west);
\draw[arrow] (colors.east)--(permutations.west);
\draw[arrow] (permutations.east)--(phi.west);
\node[note] at (4.10,3.55)
  {Bounded degree\\Inequality edges form a matching};

\node[font=\footnotesize\bfseries] at (6.10,2.90)
  {(b) The local distribution on $W$};
\node at (6.10,2.40)
  {$W\subseteq V_\phi$: every component of $J[W]$ has at most four positions};
\node[stage,text width=3.65cm,minimum width=4cm,minimum height=1.45cm]
  (common) at (2.10,1.35)
  {Common distribution of\\satisfying assignments\\of $\phi$ (YES)};
\node[stage,minimum width=1.15cm,fill=black!6]
  (local) at (6.10,1.35) {$\mu_W$};
\node[stage,text width=3.65cm,minimum width=4cm,minimum height=1.45cm]
  (computation) at (10.10,1.35)
  {Exact local computation\\of $\mu_W$ (YES and NO)};
\draw[arrow] (common.east)--node[above=5pt,note]
  {restriction\\to $W$} (local.west);
\draw[arrow] (computation.west)--(local.east);
\node[note] at (6.10,.35)
  {Every assignment of positive probability satisfies all clauses whose variables lie in $W$.};
\end{tikzpicture}
\endgroup
\label{fig:local-source}
\end{figure}
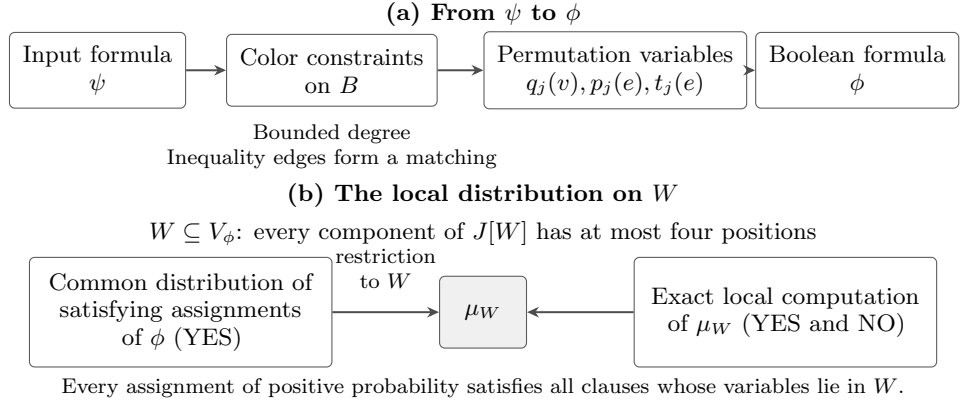

\paragraph{Color constraints.}
The input to this proposition is any constant-gap formula, including the bounded-occurrence output of Lemma~\ref{lem:pcp-source}.
We first encode the input formula by color constraints and then modify the construction so that its inequality edges form a matching.
In a satisfying coloring, a path of at most two edges then determines whether its endpoint colors are equal or different.
This property will let us compute local assignment probabilities without knowing the satisfying coloring.
Let $\psi$ have $q_0$ clauses and NO gap $\varepsilon_0$, as in the proposition.
We use a triangle on $z,\bar z,b_0$ to represent false, true, and a third color, and repeat each of its edges $q_0$ times.
For each variable $x$ with $d_x$ occurrences, we repeat every edge of the triangle on $x,\bar x,b_0$ exactly $d_x$ times.
We call such a triangle valid when its vertices have three distinct colors.
If the first triangle is valid, a valid variable triangle assigns $x$ and $\bar x$ the two Boolean colors in opposite orders.

For a clause $(t_1\lor t_2\lor t_3)$, we also use $t_1,t_2,t_3$ for its literal vertices.
We add private vertices $y,\bar y$ forming a triangle with $b_0$ and enforce $\operatorname{NAE}(t_1,t_2,y)$ and $\operatorname{NAE}(\bar y,t_3,z)$.
The not-all-equal (NAE) condition requires that the three Boolean values are not all equal.
For each such constraint, we add a triangle of three new vertices and join them to the three terminals, respectively.
If all terminals have the same Boolean color, the added triangle cannot be properly colored.
Otherwise, it admits a proper coloring.
An illustration of the triangle construction for $\operatorname{NAE}(t_1,t_2,y)$ is shown in Figure~\ref{fig:nae-gadget}.

\begin{figure}[!t]
\centering
\caption{Illustration of the 3-coloring gadget for an NAE constraint. In (a), the labels $\{1,2\}$ indicate the available colors.}
\begingroup
\linespread{1}\selectfont
\begin{tikzpicture}[x=1cm,y=1cm,
  every node/.style={font=\fontsize{9}{10.5}\selectfont},
  terminal/.style={circle,draw=black,fill=white,
    minimum size=.44cm,inner sep=1pt},
  new vertex/.style={circle,draw=black,fill=white,
    minimum size=.65cm,inner sep=1.5pt},
  triangle edge/.style={draw=black,line width=.85pt},
  terminal edge/.style={draw=black!65,line width=.7pt},
  color one/.style={fill=black!25},
  color two/.style={fill=black,text=white}]
\path[use as bounding box] (-.05,-1.95) rectangle (12.05,3.40);

\node[font=\footnotesize\bfseries] at (2.80,3.12)
  {(a) Equal terminal colors};
\node[font=\footnotesize\bfseries] at (9.20,3.12)
  {(b) Unequal terminal colors};

\foreach \shift/\panel in {0/a,6.4/b} {
  \coordinate (left\panel) at (\shift+1.50,2.10);
  \coordinate (right\panel) at (\shift+4.10,2.10);
  \coordinate (bottom\panel) at (\shift+2.80,.65);
  \coordinate (t1\panel) at (\shift+.35,2.10);
  \coordinate (t2\panel) at (\shift+5.25,2.10);
  \coordinate (y\panel) at (\shift+2.80,-.45);
  \draw[triangle edge] (left\panel)--(right\panel)--(bottom\panel)--cycle;
  \draw[terminal edge] (t1\panel)--(left\panel);
  \draw[terminal edge] (t2\panel)--(right\panel);
  \draw[terminal edge] (y\panel)--(bottom\panel);
  \node[terminal,label=above:{$t_1$}] at (t1\panel) {$0$};
  \node[terminal,label=above:{$t_2$}] at (t2\panel) {$0$};
}

\foreach \vertex in {lefta,righta,bottoma}
  \node[new vertex] at (\vertex) {$\{1,2\}$};
\node[terminal,label=below:{$y$}] at (ya) {$0$};
\node[new vertex,color one] at (leftb) {$1$};
\node[new vertex,color two] at (rightb) {$2$};
\node[new vertex] at (bottomb) {$0$};
\node[terminal,color one,label=below:{$y$}] at (yb) {$1$};

\node at (2.80,-1.15) {No proper coloring: only two colors remain.};
\node at (9.20,-1.15) {A proper coloring.};
\node[font=\fontsize{8}{9.5}\selectfont] at (6.00,-1.75)
  {Here, $0$ and $1$ are the Boolean colors, and $2$ is the third color.};
\end{tikzpicture}
\endgroup
\label{fig:nae-gadget}
\end{figure}
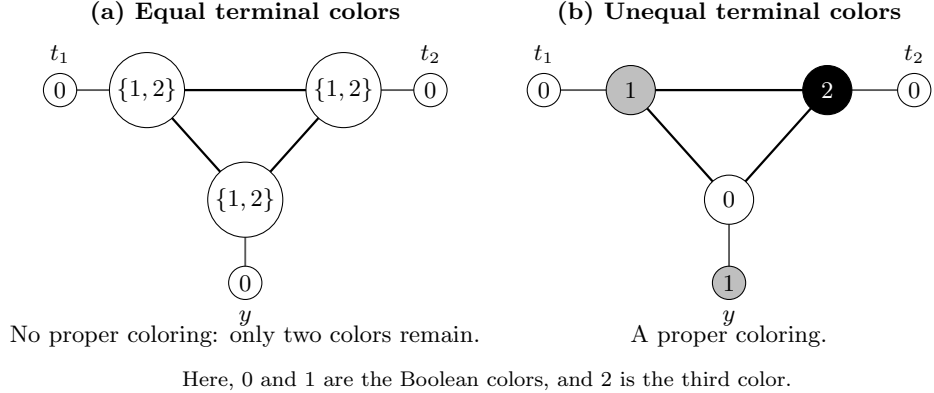

If $t_3$ is false, the second constraint forces $y$ to be false, and the first then requires $t_1$ or $t_2$ to be true.
If $t_3$ is true, we can always choose $y$ to satisfy both constraints.
Thus, the two constraints have a joint Boolean extension exactly when the clause is true.

The resulting multigraph has $\O(q_0)$ edges, since each clause has a constant-size private construction and $\sum_x d_x=3q_0$.
It is 3-colorable in YES.
In NO, an invalid triangle on $z,\bar z,b_0$ already contributes at least $q_0$ monochromatic edges.
Otherwise, we decode valid variable triangles and assign arbitrary Boolean values to variables with invalid triangles.
At least $\varepsilon_0q_0$ original clauses are false under this assignment.
We charge each false clause containing an invalid variable triangle to one such variable.
Variable $x$ receives at most $d_x$ charges, and its repeated triangle has at least $d_x$ monochromatic edges.
For each remaining false clause, the variable triangles corresponding to its literals are valid.
If its private triangle on $y,\bar y,b_0$ is invalid, that triangle has a monochromatic edge.
Otherwise, one of its two NAE constraints fails and forces a monochromatic edge in that constraint's private construction.
The charges therefore use distinct edge occurrences and account for at least $\varepsilon_0q_0$ monochromatic edges.

We next bound the degree by replacing each vertex with one copy per incident edge occurrence.
We connect these copies by equality edges from a constant-degree expander~\citep[Lemma~2.1]{journal:dinur:JACM:07}, with a fixed edge-expansion constant $\lambda>0$.
The original edges become inequality edges, and each copy belongs to exactly one of them, so these edges form a matching.
To verify the gap, consider a coloring with $f_=$ violated equality edges and $f_{\ne}$ violated inequality edges.
Within the copies of each original vertex, we recolor every copy to a most frequent color.
Each other color class has size at most half the number of copies, so expansion bounds its size by its edge boundary divided by $\lambda$.
Every violated equality edge belongs to at most two of these boundaries.
Thus, at most $2f_=/\lambda$ copies change color.
The common color in each group defines a coloring of the original multigraph.
Each changed copy affects only one inequality edge, so this coloring has at most $f_{\ne}+2f_=/\lambda$ monochromatic edges.
The preceding bound implies $\varepsilon_0q_0\le f_{\ne}+2f_=/\lambda$.

Let $B$ be the resulting graph with equality and inequality requirements.
It has constant degree, size $\Theta(q_0)$, and a constant gap.
Every equality edge joins copies of the same original vertex, whereas every inequality edge joins copies of distinct original vertices.
In particular, no pair of vertices has both an equality edge and an inequality edge.
We will use this property, together with the matching property, when comparing colors along short paths.

\paragraph{Permutation variables.}
We next use random permutations to define a distribution on satisfying assignments in YES.
The local probabilities will not depend on which satisfying coloring we start from.
To ensure this property, we use levels $0,\ldots,4$, with independent random permutations at levels $0,\ldots,3$ and fixed values at level $4$.
We will show below that four random levels suffice to make every required local distribution independent of the starting coloring.
We orient each edge $e=u\to v$ of $B$ and let $\bowtie_e$ denote its equality or inequality requirement.
Let $S_3$ be the permutations of $\lrc{0,1,2}$, with identity $1$ and composition acting from right to left.
We introduce $c_v\in\lrc{0,1,2}$ and, for $0\le j<4$, $q_j(v),p_j(e),t_j(e)\in S_3$, with $p_4(e)=1$.
For $0\le j<4$, we impose the two equalities below, together with one color constraint per edge:
\begin{equation}\label{eq:gauge-triangles}
 p_j(e)q_j(v)=t_j(e)=q_j(u)p_{j+1}(e),\qquad
 c_u\mathrel{\bowtie_e}p_0(e)c_v.
\end{equation}
Thus, each edge has eight multiplication constraints and one color constraint, each involving at most three variables.
In YES, we fix a satisfying coloring $\sigma$ of $B$, with color $\sigma_v$ at vertex $v$.
We choose independent uniform $g_{v,j}\in S_3$ for $j<4$ and set $g_{v,4}=1$.
We assign $q_j(v)=g_{v,j}g_{v,j+1}^{-1}$, $p_j(e)=g_{u,j}g_{v,j}^{-1}$, $t_j(e)=g_{u,j}g_{v,j+1}^{-1}$, and $c_v=g_{v,0}\sigma_v$.
An illustration of the permutation variables and their factors is shown in Figure~\ref{fig:permutation-constraints}.

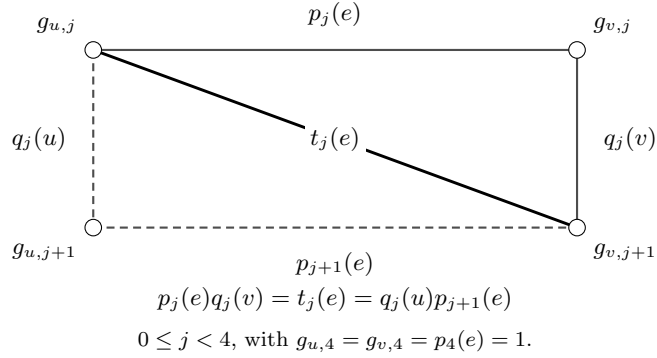
\begin{figure}[!t]
\centering
\caption{Illustration of the permutation constraints for an edge $e=u\to v$ of $B$. The two paths represent the factorizations $p_j(e)q_j(v)=t_j(e)=q_j(u)p_{j+1}(e)$.}
\begingroup
\linespread{1}\selectfont
\begin{tikzpicture}[x=1cm,y=1cm,
  every node/.style={font=\fontsize{9}{10.5}\selectfont},
  factor/.style={circle,draw=black,fill=white,inner sep=2.2pt},
  first path/.style={draw=black!70,line width=.8pt},
  second path/.style={draw=black!70,densely dashed,line width=.8pt},
  diagonal/.style={draw=black,line width=1.1pt},
  edge label/.style={fill=white,inner sep=2pt}]
\path[use as bounding box] (.20,-1.10) rectangle (10.60,3.65);

\coordinate (uj) at (2.20,3.00);
\coordinate (vj) at (8.60,3.00);
\coordinate (unext) at (2.20,.65);
\coordinate (vnext) at (8.60,.65);
\draw[first path] (uj)--node[above=5pt] {$p_j(e)$} (vj);
\draw[first path] (vj)--node[right=6pt] {$q_j(v)$} (vnext);
\draw[second path] (uj)--node[left=6pt] {$q_j(u)$} (unext);
\draw[second path] (unext)--node[below=5pt] {$p_{j+1}(e)$} (vnext);
\draw[diagonal] (uj)--node[edge label] {$t_j(e)$} (vnext);

\node[factor,label=above left:{$g_{u,j}$}] at (uj) {};
\node[factor,label=above right:{$g_{v,j}$}] at (vj) {};
\node[factor,label=below left:{$g_{u,j+1}$}] at (unext) {};
\node[factor,label=below right:{$g_{v,j+1}$}] at (vnext) {};

\node at (5.40,-.30)
  {$p_j(e)q_j(v)=t_j(e)=q_j(u)p_{j+1}(e)$};
\node[font=\fontsize{8}{9.5}\selectfont] at (5.40,-.85)
  {$0\le j<4$, with $g_{u,4}=g_{v,4}=p_4(e)=1$.};
\end{tikzpicture}
\endgroup
\label{fig:permutation-constraints}
\end{figure}

For example, $p_j(e)q_j(v)=g_{u,j}g_{v,j+1}^{-1}=q_j(u)p_{j+1}(e)$, and the color constraint follows from $\sigma_u\bowtie_e\sigma_v$.
Hence these values satisfy every constraint.

We next verify that the system retains a constant NO gap.
Consider any assignment to its variables.
For an edge whose multiplication constraints hold, the two equalities at each level imply $p_j(e)=q_j(u)p_{j+1}(e)q_j(v)^{-1}$.
We iterate this identity through the four levels, using $p_4(e)=1$, to obtain $p_0(e)=P_uP_v^{-1}$, where $P_v=q_0(v)\cdots q_3(v)$.
We decode the color at $v$ as $P_v^{-1}c_v$.
If all nine constraints associated with $e$ hold, its color constraint then implies $P_u^{-1}c_u\bowtie_e P_v^{-1}c_v$.
Thus, every violated edge of the decoded coloring forces a failed constraint associated with that edge.
Since there are only nine constraints per edge, the system retains a constant gap.
The bounded degree of $B$ also ensures that each variable occurs in only a constant number of constraints.

\paragraph{Boolean encoding.}
We now encode the color and permutation variables by Boolean variables so that the birthday graph's density bound applies.
Each Boolean position, including every auxiliary, will depend on only one color or permutation variable.
This will preserve the constant-degree property of $J$.
We first substitute the fixed value $p_4(e)=1$, so it requires no Boolean variables.
For a color or permutation variable $X$, we use an indicator $[X=a]$ for each of its three or six possible values $a$.
We require $\bigvee_a[X=a]$ and $\neg[X=a]\lor\neg[X=b]$ whenever $a\ne b$, so a valid code specifies exactly one value.

Each constraint has its own block of clauses enforcing valid codes for its inputs and excluding its rejected input tuples.
If its inputs are $X_1,\ldots,X_t$, where $t\le3$, a rejected tuple $(a_1,\ldots,a_t)$ contributes the clause $\bigvee_{i=1}^t\neg[X_i=a_i]$.
Only the clause requiring one of six indicators to be true has more than three literals.
We replace $z_1\lor\cdots\lor z_6$ by $(z_1\lor z_2\lor y_1)$, $(\neg y_1\lor z_3\lor y_2)$, $(\neg y_2\lor z_4\lor y_3)$, and $(\neg y_3\lor z_5\lor z_6)$.
For a valid code, these clauses hold with $y_j=\bigwedge_{i=1}^{j+1}\neg z_i$ for $j=1,2,3$.
We use these auxiliary values, each determined by the single variable encoded by $z_1,\ldots,z_6$.
For an all-zero code, the first three clauses force $y_1=y_2=y_3=1$, which falsifies the last clause.

Different constraint blocks share only the indicators $[X=a]$, while their clause occurrences and auxiliary variables are separate.
If a block is satisfied, every input code is valid and the encoded constraint holds.
To prove the gap, decode valid codes and choose arbitrary values for invalid codes.
Any constraint that fails after this step must already have a false clause in its own block, since a satisfied block would have decoded correctly.
We choose one such clause occurrence for each failed constraint.
The chosen occurrences are distinct because different blocks have separate clause occurrences.
In NO, the decoded assignment violates a constant fraction of the constraints, and each block contains only a constant number of clauses.
Thus, a constant fraction of the clauses of $\phi$ are false, so $\phi$ has a constant NO gap $\varepsilon_1>0$ depending only on $\varepsilon_0$.
It has $s=\Theta(q_0)$ variables and $\O(q_0)$ clauses, and every variable has constant occurrence.
In YES, the satisfying values above, their indicator codes, and the specified auxiliary values satisfy $\phi$.
This defines one common distribution of satisfying Boolean assignments.
Short clauses may repeat a literal if exactly three literals are required.

\paragraph{Local distributions.}
We finally construct $J$ and compute $\mu_W$ without knowing the satisfying coloring $\sigma$.
Associate every Boolean position, including each auxiliary, with its one color or permutation variable.
We join two distinct positions in $J$ exactly when their associated variables use a common random permutation $g_{v,j}$ with $j<4$.
We use a separate graph to track how many colors can affect one component of $J[W]$, with vertex $(v,j)$ for the factor $g_{v,j}$.
The variable $q_j(v)$ uses the edge from $(v,j)$ to $(v,j+1)$, $p_j(e)$ uses the edge from $(u,j)$ to $(v,j)$, and $t_j(e)$ uses the edge from $(u,j)$ to $(v,j+1)$.
The variable $c_v$ uses $g_{v,0}$, and all level-four values are fixed.
Since $B$ has bounded degree, each random $g_{v,j}$ occurs in only a constant number of these variables.
Each variable has only a constant number of indicators and auxiliaries, so $J$ has constant degree.

Fix $W\subseteq V_\phi$ such that every component of $J[W]$ has at most four positions.
In one component, let $r$ count its distinct associated color variables and let $b$ count its distinct associated permutation variables.
Each variable has at least one position in the component, so $r+b\le4$.
Each associated color variable corresponds to a level-zero vertex in the graph above, and each associated permutation variable corresponds to an edge.
Adjacent positions share a random permutation, so the vertices and edges associated with the positions in this component form a connected subgraph.
Its $b$ edges therefore connect the $r$ distinct level-zero color vertices, which requires $b\ge r-1$.
Together with $r+b\le4$, this implies $2r-1\le4$ and hence $r\le2$.
If $r\ge1$, then $b\le3$.
Each permutation edge changes the level by at most one, so these edges cannot connect a level-zero color to level four.

If $r=2$, then $b\le2$, so a path between the colors uses at most two permutation edges.
After we ignore the level indices and omit steps that stay at the same vertex of $B$, this path uses at most two edges of $B$.
Its endpoints are distinct color variables, so the path cannot traverse the same edge twice.
Since the inequality edges form a matching, at most one edge on this path is an inequality edge.
Thus, every satisfying coloring assigns equal colors to the endpoints if the path contains no inequality edge, and different colors otherwise.
With at most two permutation edges, different paths between the colors can only be parallel edges.
The construction of $B$ excludes parallel edges of opposite types, so the resulting equal-or-different relation is unambiguous even in NO.

We can now compute the local distribution.
For a component with one color variable $c_v$, we replace $\sigma_v$ by $0$ in the expression $c_v=g_{v,0}\sigma_v$.
For two color variables, we replace their underlying colors by $(0,0)$ if their relation is equality and $(0,1)$ otherwise.
For a component without colors, use the permutation expressions directly.
Distinct components share no random permutations, so we can sample them independently.
We enumerate the at most $2\size{W}$ random permutations used by the positions in $W$, with level-four values fixed to $1$.
For every choice, we evaluate the associated variables, their indicator codes, and their specified auxiliary values on $W$.
Their frequencies define $\mu_W$ and have $\O(\size{W})$-bit numerators and denominators.
Thus, we can list the assignments of positive probability and their exact probabilities in $6^{\O(\size{W})}s^{\O(1)}$ time.
The procedure uses only the constructed constraints and requires no satisfying coloring.

In YES, we compare this distribution with the restriction to $W$ of the common distribution defined above.
For a component containing colors, choose a permutation $\pi$ taking their underlying colors in $\sigma$ to the fixed values just specified.
Such a permutation exists because their equal-or-different relation is the one forced by the path in $B$.
Within this component, replace every used $g$ by $g\pi^{-1}$ and each underlying color by its image under $\pi$.
Every permutation product remains unchanged, as does each encoded color, since $(g\pi^{-1})(\pi\sigma_v)=g\sigma_v$.
The component uses no fixed level-four value, so this change preserves the independent uniform distribution of its random permutations.
The computed component distribution therefore equals the corresponding restriction of the common YES distribution.
A component without colors already uses the same permutation expressions.
Independence across components proves the required equality of distributions on $W$.

It remains to verify that every assignment of positive $\mu_W$-probability satisfies all clauses contained in $W$, including in NO.
For a clause that enforces a valid code, the positions in $W$ receive their values from the same complete valid code and the auxiliary values specified above.
Hence the clause holds, even if other positions of that code lie outside $W$.
Every other clause excludes a rejected tuple of one constraint and contains an indicator of each of its nonconstant inputs.
For a multiplication constraint, these input expressions form a triangle of permutation products, or a two-variable equality after the fixed value $p_4(e)=1$ is substituted.
They are connected through shared random permutations, so all their indicators in the clause lie in one component of $J[W]$.
The procedure therefore evaluates the inputs with the same random permutations, and the multiplication identity holds.
For a color constraint, the indicators of $c_u,p_0(e),c_v$ in the clause similarly lie in one component.
Its fixed underlying colors satisfy the equality or inequality required by $e$, so their transformed values satisfy $c_u\bowtie_e p_0(e)c_v$.
Thus, the tuple clauses also hold.
The argument uses an indicator of every input but does not require its whole code to lie in $W$, which proves local satisfaction for every clause contained in $W$.
\end{proof}

\subsection{Proof of Lemma~\ref{lem:retained-mass}}\label{app:retained-mass}
\begin{proof}
A pair is good if and only if $J[S\cup T]$ contains no connected set of five positions.
We first analyze a uniform $2\ell$-set, because $S\cup T$ contains at most $2\ell$ positions.
For $0\le j\le s$, let $b_j$ be the probability that every component of $J[W]$ has at most four positions when $W\subseteq V_\phi$ is a uniform $j$-set.
If we remove a uniformly chosen position from a uniform $(j+1)$-set, the result is a uniform $j$-set.
Every remaining component lies in an original component, so $b_j\ge b_{j+1}$.

We now include each position independently with probability $p=4\ell/s$ to obtain a random set $X$.
We may assume $g_0\ge8$, so $p\le1/2$ and $2\ell\le s$.
Conditional on $\size{X}=j$, the set $X$ is a uniform $j$-subset.
By the law of total probability, $\EE{b_{\size{X}}}$ is therefore the probability that every component of $J[X]$ has at most four positions.
Every connected set of five positions admits an ordering in which each position after the first has an earlier neighbor.
There are at most $s,D,2D,3D,4D$ choices at successive steps, so there are at most $24sD^4$ such sets.
Every component of $J[X]$ has at most four positions exactly when $X$ contains none of these sets.
For each such five-set, the event that it is not contained in $X$ is decreasing, since this property is preserved when positions are removed.
By repeatedly applying Harris's inequality~\citep[Theorem~6.3.2]{book:alon:BOOK:08}, the probability that all these events hold is at least the product of their probabilities. Consequently,
\[
\EE{b_{\size{X}}}
\ge (1-p^5)^{24sD^4}
\ge e^{-E},
\qquad E=C_D\frac{\ell^5}{s^4},
\]
for a sufficiently large constant $C_D>0$
depending only on $D$.
For the second inequality, we use $\ln(1-x)\ge-2x$ with $x=p^5\le1/2$.

Since $\EE{\size{X}}=4\ell$, the Chernoff bound implies $\Pr(\size{X}<2\ell)\le e^{-\ell/2}$.
On $\size{X}\ge2\ell$, monotonicity implies $b_{\size{X}}\le b_{2\ell}$, while $b_{\size{X}}\le1$ always.
Hence, we have
\[
 e^{-E}\le\EE{b_{\size{X}}}
 \le b_{2\ell}+\Pr(\size{X}<2\ell)
 \le b_{2\ell}+e^{-\ell/2}.
\]
Take $g_0$ large enough that $E/\ell\le C_D/g_0^4\le1/4$.
Since $\ell\ge s^{7/8}/2$, for sufficiently large $s$, we have $e^{E-\ell/2}\le e^{-\ell/4}\le1/2$, and hence $b_{2\ell}\ge\frac12e^{-E}$.

Finally, choose an unordered pair $\lrc{S,T}$ uniformly among all pairs of distinct $\ell$-sets.
Conditional on $\size{S\cup T}=j$, the union is a uniform $j$-set by permutation symmetry.
Since $j\le2\ell$ and $b_j$ is nonincreasing,
\[
\zeta=\EE{b_{\size{S\cup T}}}
\ge b_{2\ell}\ge\frac12e^{-E},
\]
which proves the claimed bound.
\end{proof}

\section*{Acknowledgement}
ChatGPT was used to assist with research exploration, proof checking, and manuscript preparation. The authors verified the results and take full responsibility for the content of this paper.

\bibliographystyle{apalike}
\bibliography{DaRP}
\end{document}